\documentclass[10pt,prb,twocolumn,aps,amsmath,amssymb,showpacs]{revtex4-1}

\usepackage{graphicx}
\usepackage{dcolumn}
\usepackage{bm}
\usepackage{wrapfig}
\usepackage{amsmath,amsfonts,amssymb,bm}
\usepackage[usenames]{color}
\usepackage[title]{appendix}

\begin{document}


\title{Thermodynamics of a Quantum Ising system coupled to a Spin Bath: Zero Temperature Results}

\author{ R.D. McKenzie$^{1}$, P.C.E. Stamp$^{1,2}$ }

\affiliation{$^1$ Department of Physics
and Astronomy, University of British Columbia, Vancouver B.C.,
Canada V6T 1Z1. \\
$^2$ Pacific Institute of Theoretical Physics, University of British Columbia, Vancouver B.C.,
Canada V6T 1Z1.
 }


\begin{abstract}
We study the effect of coupling a spin bath environment to a system which, at low energies, can be modeled as a quantum Ising system. A field theoretic formalism incorporating both thermal and quantum fluctuations is developed to derive results for the thermodynamic properties and response functions, both for a toy model and for the $LiHoF_4$ system, in which spin-8 electronic spins couple to a spin-$7/2$ nuclear spin bath: the phase transition then occurs in a system of electronuclear degrees of freedom, coupled by long-range dipolar interactions. The quantum Ising phase transition still exists, and one hybridized mode of the Ising and bath spins always goes soft at the transition.

\end{abstract}


\maketitle


\section{Introduction}
 \label{sec:intro}
 

In both statistical physics and quantum computation the Quantum Ising model plays a central role \cite{SuzukiBook, DuttaBook}. It is key to understanding quantum phase transitions \cite{SachdevBook} (QPTs), and describes a large variety of solid-state and atomic spin systems \cite{WolfIsing, KimSimulations, BlattSimulations}, as well as many quantum computational systems \cite{KadowakiNishimori,Farhi2001,SantoroSimulations,JohnsonRoseSCQubits}.

Although the Quantum Ising model has been studied extensively over the years,
one key unanswered question does stand out, viz., what is the effect of a coupling to its environment? This problem is not only of theoretical interest; it also has large practical ramifications. Both the thermodynamics and dynamics of a large variety of Quantum Ising systems - ranging from quantum information processing systems to magnetic, superconducting, and atomic spin ensembles - are affected by their environments. When one turns to systems being devised for quantum computation, the mechanisms governing decoherence must be understood if one is to have any hope of making them work.

The 1-dimensional quantum Ising model coupled to an ``oscillator bath" environment \cite{FeynmanVernon} (modelling extended degrees of freedom like phonons or photons) has received considerable attention \cite{[{}][{ In particular, see Section 9.2 and references therein.}] CarrBook, HoyosSmearedQPT}, and the bath has been shown to have a significant impact on the quantum critical behaviour of the model. We will take the view here that much of this physics is reasonably well understood.

However, the result of coupling a ``spin bath" environment \cite{ProkofievStampRPP} (modelling spatially localised degrees of freedom like nuclear and paramagnetic spins, and various solid-state defects) to a Quantum Ising system is not so clear. Experiments on
$LiHoF_4$, often considered the archetypal solid-state Quantum Ising system, have suggested \cite{RonnowScience05, Ronnow07} that the mode softening expected at the QPT is suppressed by coupling to a spin bath environment; very recent experiments \cite{KovacevicRonnow16} have probed transitions between electronuclear modes \cite{SchechterStampPRL} in $LiHoF_4$.
Spin bath modes also cause strong decoherence \cite{ProkofievStampRPP,YaoSham06, [{}] [{ For recent work see }] MartinisSQUID, *Sarabi2LevelDefects, *Lisenfeld2LevelDefects, Takahashi11}. In adiabatic quantum computation \cite{KadowakiNishimori,Farhi2001,SantoroSimulations,JohnsonRoseSCQubits}, suppression of the QPT would be expected to radically change the dynamics. Recently, it has been proposed that an AC field may be used to control the strength of the couplings between the Ising and bath spins \cite{GomezLeonStamp}, opening up a rich testing ground for quantum critical behaviour. Thus a lot turns on the question of how a spin bath affects a Quantum Ising system.

There are a number of ways one can approach this problem. One is to try and set up a ``theoretical minimum" toy model, which captures all the essential physics without becoming too complicated.  Another is to look at a real experimental system, such as the $LiHoF_4$ system, for which extensive data exists, and where one can reasonably hope to make accurate and testable theoretical predictions.

In the present paper we develop both a toy model and a detailed model for $LiHoF_4$, and address the physics surrounding the QPT for both of them. The results are thus not only useful in understanding $LiHoF_4$; they also give us a good understanding of what is essential (and what is not essential) in any model. 

We emphasize that the study here focuses on the physics of the QPT, and thus does not deal with several other important questions. In particular, we only develop the toy model to the point where we can extract conclusions about the QPT - a more detailed analysis will appear in another paper. We also restrict the study in this paper to the case of temperature $T=0$ (again, because we are focussing on the QPT); the finite $T$ case will also be dealt with elsewhere. 

The approach we use is fairly conventional, and goes back to old work on quantum phase transitions \cite{Hertz, YoungQPT, Millis93}: first an auxiliary field is introduced representing order parameter fluctuations, then a trace is performed over all degrees of freedom (the Ising and bath spins) apart from those associated with the ordering field. The resulting effective scalar field theory accounts for both quantum and thermal fluctuations in the underlying microscopic Hamiltonian. The primary difference between our work and previous work is that the trace is performed over the Ising and bath spins rather than, for example, itinerant fermions. We find that even though a gap forms in the Ising mode spectrum, a new hybridized mode between the Ising and bath spins appears, which fully softens at the QPT. In many solid-state spin systems this will be an electronuclear mode; we discuss various experiments for probing this mode, including the NMR used in recent experiments \cite{KovacevicRonnow16} on $LiHoF_4$.

The paper is organized as follows. In Section \ref{sec:effH} we introduce the Toy Model Hamiltonian, and then derive the low-energy effective Hamiltonian for the $LiHoF_4$ system, in terms of known parameters coming from a more microscopic Hamiltonian; these are the models we work with in the rest of the paper. Once this is in done, we develop the field-theoretic techniques required to calculate the properties of these Hamiltonians, in Section \ref{sec:FormToy}. These are then used in the Random Phase Approximation (RPA) in Section \ref{sec:CorrFuncs} to derive the dynamic susceptibilities for the two models; in Section \ref{sec:Qfluc} we go beyond the RPA, adding the effect of quantum fluctuations up to 4th order in the fluctuation fields; we calculate the phase diagram and the magnetization and show that the QPT survives the coupling to the spin bath. Finally, in Section \ref{sec:Exp} we apply the results to understand experiments in several systems. The calculation of formulas for the magnetization and the cumulants of the partition function is lengthy and gives quite complex expressions - these calculations are relegated to two Appendices.


\section{effective Hamiltonians}
\label{sec:effH}


The $LiHoF_4$ system is a lattice of $Ho$ ions, with each $Ho$ surrounded by a cage of $Li$ and $F$ ions. The effective Hamiltonian usually used to describe this system is defined in terms of the net magnetic moment of the electrons of each $Ho$ ion and their interactions with each other and its nuclear moment. The effect of the $Li$ and $F$ ions is incorporated via the inclusion of a crystal electric field. In what follows we we give the full details of this Hamniltonian, and then show how it can be truncated to a much simpler model, valid at temperatures well below $10K$, in which the spin-$8$ Ho ions are truncated to a lattice of 2-level systems. Because the details are complicated, we first briefly recall, as a kind of baseline, the form of the simple ``toy model" referred to in the introduction.

\subsection{Toy Model Hamiltonian}
\label{sec:toyM}

The Quantum Ising model on its own is defined by one of the simplest Hamiltonians in physics; in terms of Pauli operators $\{ \mbox{\boldmath $\tau$}_j \}$, it is written as
\begin{equation}
\mathcal{H}^{toy}_0  =  -\sum_{i<j}V_{ij}\ \tau_{i}^{z}\tau_{j}^{z}-\Delta_0 \sum_{i} \tau_{i}^{x}
 \label{eq:H1}
\end{equation}
where each two level system feels a transverse tunneling field $\Delta_0$, and is coupled to its neighbours via the longitudinal interactions $V_{ij}$. The competition between these two terms causes a QPT \cite{SachdevBook} between an ordered phase for $g = |\Delta_0/V_0| < g_c$ (where $V_0 = \sum_j V_{ij}$ and $g_c \sim O(1)$ in the absence of the spin bath $\mathcal{H}_{\textrm{SB}}$), and a disordered phase for $g > g_c$. In adiabatic quantum computation and quantum annealing \cite{KadowakiNishimori,Farhi2001}, the parameters $\Delta_0$ and $V_{ij}$ are varied slowly in time.

Both ``oscillator bath" modes \cite{FeynmanVernon} and ``spin bath" modes \cite{ProkofievStampRPP} can couple to the Ising spins. To capture the essential effects of coupling to a spin bath, one assumes a lattice of central ``Ising" spins $\{ \mbox{\boldmath $\tau$}_j \}$ couples locally to a set of two level ``bath" spins $\{ \mbox{\boldmath $\sigma$}_j \}$ (so that on each lattice site $j$ we have spin pair states $| \mbox{\boldmath $\tau$}_j, \mbox{\boldmath $\sigma$}_j \rangle$). We can write this spin bath term as
\begin{equation}
\mathcal{H}^{toy}_{\textrm{SB}} \;=\; A_{z} \sum_{i} \sigma_i^{z} \tau_i^{z} + \frac{A_{\perp}}{2} \sum_i (\sigma_i^{+} \tau_i^{-} + \sigma_i^{-} \tau_i^{+})
 \label{SH2-hyp}
\end{equation}
having both longitudinal and transverse interactions.

One can easily add to this spin bath coupling a set of couplings to harmonic oscillators, of the standard ``spin-boson" form \cite{LeggettSpinBoson87}, representing phonons (as well as photons, if necessary). In this paper we will ignore these bosonic bath modes, since our primary concern is the effect of the spin bath.
Thus our toy model will be described by the effective Hamiltonian
\begin{equation}
{\cal H}^{toy} = {\cal H}^{toy}_0 + {\cal H}^{toy}_{SB}
 \label{Htoy}
\end{equation}
and in what follows we will from time to time compare its behaviour with the predictions we make for the $LiHoF_4$ system.

\subsection{The $LiHoF_4$ System}
\label{sec:LiHo}

We consider the classic 3-dimensional Quantum Ising magnet $LiHoF_4$. This material has subtle (and sometimes controversial) experimental properties \cite{Rosenbaum,JonssonBarbara,QuilliamKyciaPRL07,RodriguezLukeMuSRLiHo}, many of which clearly depend on the coupling to its spin bath environment \cite{RonnowScience05, QuilliamKyciaPRL07,SchmidtAeppli14}. It differs in four key ways from the toy model, viz. (i) the Quantum Ising spins result from truncation of spin-$8$ ionic spins $\{ {\bf J}_j \}$; (ii) the bath is now made up of nuclear spins $\{ {\bf I}_j \}$, with spin-$7/2$, not spin-$1/2$; (iii) in a transverse applied field, the hyperfine coupling actually generates a transverse term acting directly on the bath spins, absent from our toy model; (iv) the inter-spin Ho-Ho interactions are now long-range dipolar. Clearly any one of these features might render the toy model conclusions invalid; thus, if we are to believe that the toy model results are in any way generic, we must generalize the previous discussion to include all these extra features.

The total ``microscopic" Hamiltonian for $LiHoF_4$ is given by \cite{ChakrabortyGirvin}
\begin{align}
\label{eq:LiHoHam}
\mathcal{H} = &\sum_{i} V_{C}(\vec{J_{i}}) - g_{L}\mu_{B}\sum_{i}B_{x}J_{i}^{x}
+A\sum_{i} \vec{I_{i}} \cdot \vec{J_{i}}
\\ \nonumber
&- \frac{1}{2} J_{D} \sum_{i \neq j} D_{ij}^{\mu\nu} J_{i}^{\mu}J_{j}^{\nu}
+ \frac{1}{2} J_{nn} \sum_{ <ij>} \vec{J_{i}} \cdot \vec{J_{j}},
\end{align}
where by ``microscopic" one implies, as usual in quantum magnetism, that the energy scale assumed is to be well below that where one needs to get into the internal atomic physics of individual ions. Thus $V_{C}(\vec{J_{i}})$ is the crystal electric field energy, $B_{x}$ is an applied transverse magnetic field, $D_{ij}^{\mu\nu}$ is a dipolar interaction between electronic spins with $J_{D} = \frac{\mu_{0}}{4\pi} (g_{L}\mu_{B})^{2}$, the antiferromagnetic exchange interaction is $J_{nn}=1.16mK$, and $A=39mK$ is the hyperfine interaction. We have electronic spin $J=8$ and nuclear spin $I=\frac{7}{2}$. The Land{\'e} $g$ factor is $g_{L} = \frac{5}{4}$, and $\mu_{B} = 0.6717K/T$ is the Bohr magneton.  

The site summations in (\ref{eq:LiHoHam}) are over a tetragonal Bravais lattice with four $Ho^{3+}$ ions per unit cell. The lattice spacing in the $xy$ plane is $a=5.175$ Angstroms and the longitudinal lattice spacing is $c=10.75$ Angstroms.  The holmium ions have fractional coordinates $(0,0,\frac{1}{2})$, $(0,\frac{1}{2},\frac{3}{4})$, $(\frac{1}{2},\frac{1}{2},0)$ and $(\frac{1}{2},0,\frac{1}{4})$. We neglect quadrupole interactions because they are small, and ignore here the nuclear spins on the $F$ and $Li$ sites, since their hyperfine couplings to the $Ho$ electronic spins are too weak to have an affect on the thermodynamic properties.

We now wish to truncate this microscopic Hamiltonian (\ref{eq:LiHoHam}) down to an effective Hamiltonian in a 16 dimensional subspace (per site). We do this by (i) determining effective spin- 1/2 operators for the electronic spins by truncating the terms in the spin-8 single-ion electronic component of the Hamiltonian down to a $2 \times 2$ subspace; and (ii) applying the truncation to the hyperfine component of the full microscopic Hamiltonian to obtain our final effective Hamiltonian. We will truncate, in turn, the electronic and nuclear terms.

\subsubsection{Truncation of the Electronic Terms}
 \label{sec:elecTrunc}

The single-ion Hamiltonian for the spin-$8$ electronic component of the $Ho$ ions is
\begin{align}
\mathcal{H}_{0e} = & V_{C}(\vec{J}) - g_{L}\mu_{B} B_{x}J^{x}.
\end{align}
The crystal field Hamiltonian $V_{C}(\vec{J})$ has the form
\begin{align}
V_{C}(\vec{J}) &= B_{2}^{0}O_{2}^{0} + B_{4}^{0}O_{4}^{0} + B_{6}^{0}O_{6}^{0}
+ B_{4}^{4}(C)O_{4}^{4}(C)
\\ \nonumber
&+ B_{6}^{4}(C)O_{6}^{4}(C) + B_{4}^{4}(S)O_{4}^{4}(S) + B_{6}^{4}(S)O_{6}^{4}(S),
\end{align}
where we use the standard Stevens' operators $O_n^m$, and for the $B_n^m$ we use the estimates of R{\o}nnow \textit{et al} \cite{Ronnow07}. The eigenstates of the crystal field are mixed and split by an applied transverse field $B_x$.

\begin{figure}[htp]
\centering
\includegraphics[width=8.5cm]{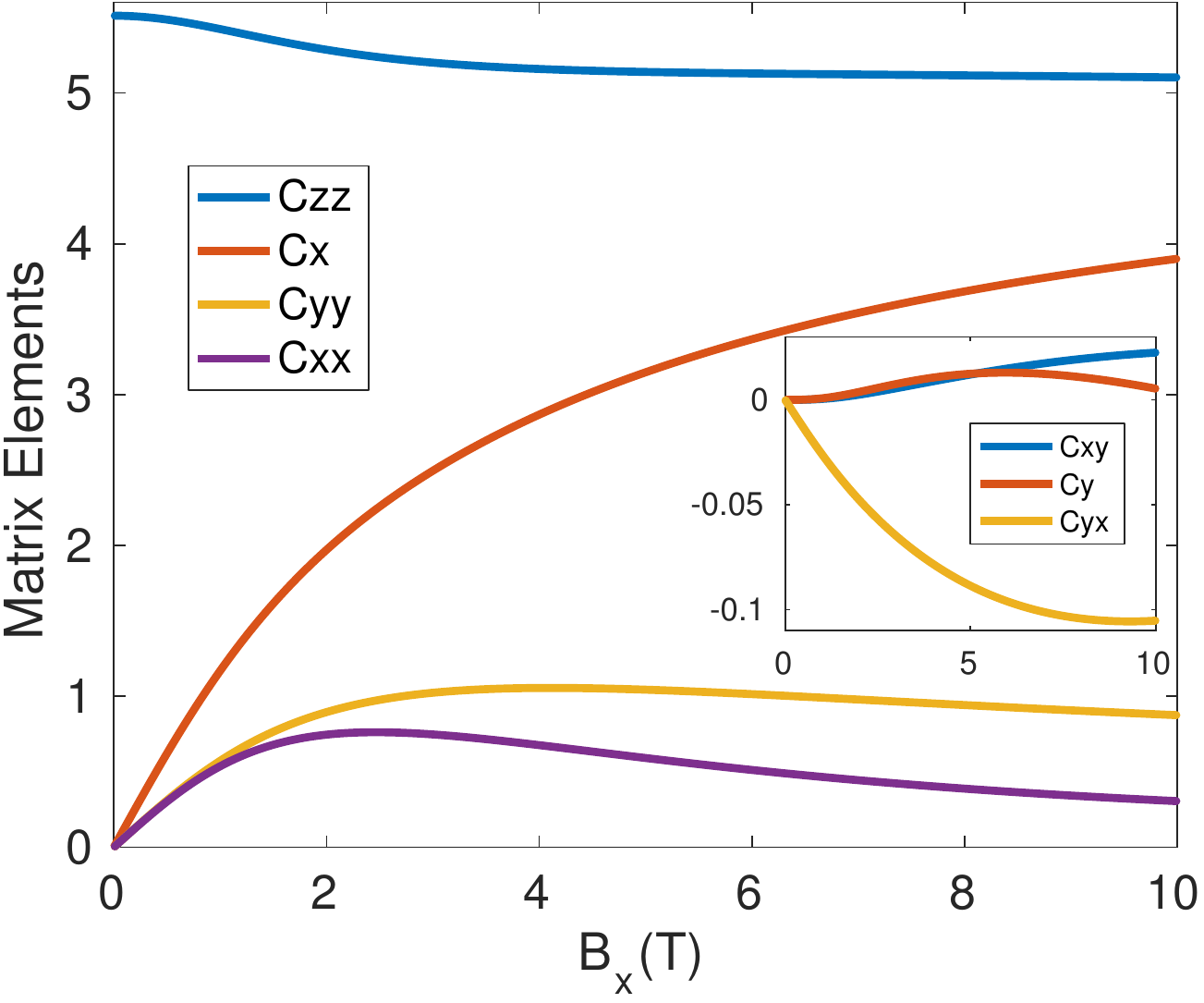}
\caption{The non-zero matrix elements of the effective spin half operators, $J^{\mu} = C_{\mu}(B_x) + \sum_{\nu=x,y,z} C_{\mu\nu}(B_{x})\tau^{\nu}$, for the truncated $LiHoF_4$ Hamiltonian, as a function of the applied transverse field $B_x$ (in Tesla).}
\label{fig:ME}
\end{figure}

Following the procedure of Chakraborty \textit{et al} \cite{ChakrabortyGirvin}, we now diagonalize the electronic single ion Hamiltonian $\mathcal{H}_{0e}$, using a unitary rotation $U$, such that  $\mathcal{H}_{0e} \rightarrow \widetilde{\mathcal{H}}_{0e} = U \mathcal{H}_{0e} U^{\dagger}$, and $J^{\mu} \rightarrow \widetilde{J}^{\mu} = U J^{\mu} U^{\dagger}$. We then truncate the operators down to the two-dimensional subspace involving the two lowest eigenstates of $\mathcal{H}_{0e}$; the original spin operators $J^{\mu}$ may then be expressed in terms of Pauli operators $\tau^{\mu}$ operating on the $2 \times 2$ subspace in the form
\begin{align}
J^{\mu} = C_{\mu}(B_x) + \sum_{\nu=x,y,z} C_{\mu\nu}(B_{x})\tau^{\nu}.
\end{align}
The lower two electronic eigenstates of $\mathcal{H}_{0e}$ are separated from the rest of the electronic eigenstates by a gap of at least $10.3K$. The hyperfine interaction, and the interactions between holmium spins, are too weak to cause significant mixing with the higher lying eigenstates, which justifies the truncation procedure. We apply a second rotation in order to diagonalize the $J^z$ operator in the $2 \times 2$ subspace so that $J^{z}=C_{zz}\tau^{z}$. In terms of the two lowest eigenstates of $\mathcal{H}_{0e}$, $|\alpha\rangle$ and $|\beta\rangle$, our basis is $|\uparrow\rangle = \frac{1}{\sqrt{2}}[|\alpha\rangle + \exp{i\theta}|\beta\rangle]$, and $|\downarrow\rangle = \frac{1}{\sqrt{2}}[|\alpha\rangle - \exp{i\theta}|\beta\rangle]$, where the phase is fixed such that the coefficient of the lowest eigenstate $|\alpha\rangle$ is real and positive. In Fig. (\ref{fig:ME}), we plot the non-zero matrix elements of the effective spin half operators as a function of the transverse field.


The interactions between electronic spins must also be truncated. Applying the truncation procedure to the electronic spins in (\ref{eq:LiHoHam}) we find $\mathcal{H}_{int} = -\frac{1}{2}\sum_{\bf k} V_{\bf k} \tau_{\bf k}^z \tau_{\bf -k}^z$, with
\begin{equation}
\label{eq:V-dip}
V_{\bf k} = C_{zz}^2 [J_{D} D_{\bf k}^{zz}-J_{nn}\gamma_{\bf k} ],
\end{equation}
where $D_{\bf k}^{zz}$ is the shape dependent Fourier transform of the dipolar interaction, and
\begin{align}
\gamma_{\bf k} = 2\cos{\biggr(\frac{k_z c}{4}\biggr)}\biggr[\cos{\biggr(\frac{k_x a}{2}\biggr)}
+\cos{\biggr(\frac{k_y a}{2}\biggr)}\biggr]
\end{align}
is the Fourier transform of the exchange interaction, incorporating the four nearest neighbour atoms at $(\pm \frac{a}{2},0,-\frac{c}{4})$ and $(0,\pm \frac{a}{2}, \frac{c}{4})$. The electronic dipole-dipole interaction is strongly anisotropic; the physical source of this anisotropy is the deformation of the electronic $4f$ orbitals due to the crystal electric field. The antiferromagnetic exchange interaction $J_{nn}$ between nearest neighbor sites is much weaker than the dipole-dipole interaction; we use the estimate of R{\o}nnow \textit{et al}  \cite{Ronnow07}, $J_{nn} = 1.16mK$. In a long cylindrical sample of $LiHoF_4$ the strength of the dipolar interaction at zero wavevector is $J_D D_0^{zz} = 78.9mK$, which should be compared to the exchange energy $J_{nn} \gamma_0 = 4.64mK$.

In terms of the effective spin operators, the total electronic Hamiltonian $\mathcal{H}_e$ may now be written in terms of Pauli operators in the $2 \times 2$ subspace as
\begin{align}
\label{eq:Hetrunc}
\mathcal{H}_e \approx  - \frac{1}{2}\Delta(B_{x})&\sum_{i}\tau_{i}^{x}
- \frac{1}{2}J_{D}C_{zz}^2(B_x) \sum_{i \neq j} D_{ij}^{zz} \tau_{i}^{z}\tau_{j}^{z}
\\ \nonumber
& + \frac{1}{2} J_{nn} C_{zz}^2(B_x) \sum_{ <ij>} \tau_{i}^{z} \tau_{j}^{z}.
\end{align}
The terms neglected in this approximation either vanish due to symmetry considerations, or they are significantly smaller ($\sim 1\%$) than the terms given in equation (\ref{eq:Hetrunc}) (for a discussion of these correction terms see Tabei \textit{et al}\cite{Tabei}).

The Ising nature of the system is now apparent; indeed, we can rewrite ${\cal H}_e$ in the toy model form (\ref{eq:H1}) as
\begin{align}
\label{eq:Hetrunc1}
\mathcal{H}_e \approx =  -\sum_{i<j}V_{ij}(B_x)\ \tau_{i}^{z}\tau_{j}^{z}-\Delta (B_x) \sum_{i} \tau_{i}^{x}
\end{align}
with the parameter $V_{ij}(B_x)$ given by
\begin{equation}
V_{ij} \;=\; 
\frac{1}{2} \bigr(J_{D} D_{ij}^{zz}C_{zz}^2(B_x) - J_{nn} \delta_{ij} C_{zz}^2(B_x)\bigr)
 \label{eq:Hetrunc2}
\end{equation}
so that both the ``tunneling term" $\Delta(B_x)$ and the Ising interaction $V_{ij}(B_x)$ now have a very pronounced dependence on the applied field $B_x$.

\subsubsection{Truncation of the Nuclear Spin Terms}
 \label{sec:nuclTrunc}

We now reintroduce the nuclear spins by truncating the hyperfine interaction, $\mathcal{H}_{hyp} = A\sum_{i} \vec{I_{i}} \cdot \vec{J_{i}}$, down to the lowest two electronic levels (we replace $\vec{J_{i}}$ with the effective spin half operator for the $2 \times 2$ subspace). This is done by applying the same truncation procedure used above for the purely electronic component of the Hamiltonian. Keeping only non-zero terms, the result is then (suppressing the field dependence $B_x$ of all operators):
\begin{align}
\label{eq:EffectiveHamiltonian}
\mathcal{H}_{NS} =
  &\sum_{i} \vec{\Delta}_{n} \cdot {\bf I}_{i} + A_{z} \sum_{i} \tau_{i}^{z}I_{i}^{z}
\\ \nonumber
&+ \left( A_{\perp} \sum_{i} \tau_{i}^{+}I_{i}^{-}
 + A_{++} \sum_{i} \tau_{i}^{+}I_{i}^{+}
+  h.c. \right)
\end{align}
where
\begin{align}
\vec{\Delta}_{n} = (A C_{x},A C_{y},0)
&&
A_{z} &= A C_{zz},
\end{align}
and
\begin{align}
A_{\perp} &= A \frac{C_{xx}+C_{yy}+i(C_{yx}-C_{xy})}{4}
\\ \nonumber
A_{++} &= A \frac{C_{xx}-C_{yy}-i(C_{yx}+C_{xy})}{4}.
\end{align}
The effective field $\vec{\Delta}_{n}$ is a result of the strong hyperfine interaction in $LiHoF_4$; the physical transverse field shifts the electronic $4f$ orbitals, leading to a static effective field. Thus we end up with a nuclear dynamics governed both by this static field, and by the time-varying fields coming from the electronic spins.

\begin{figure}[htp]
\centering
\includegraphics[width=8.5cm]{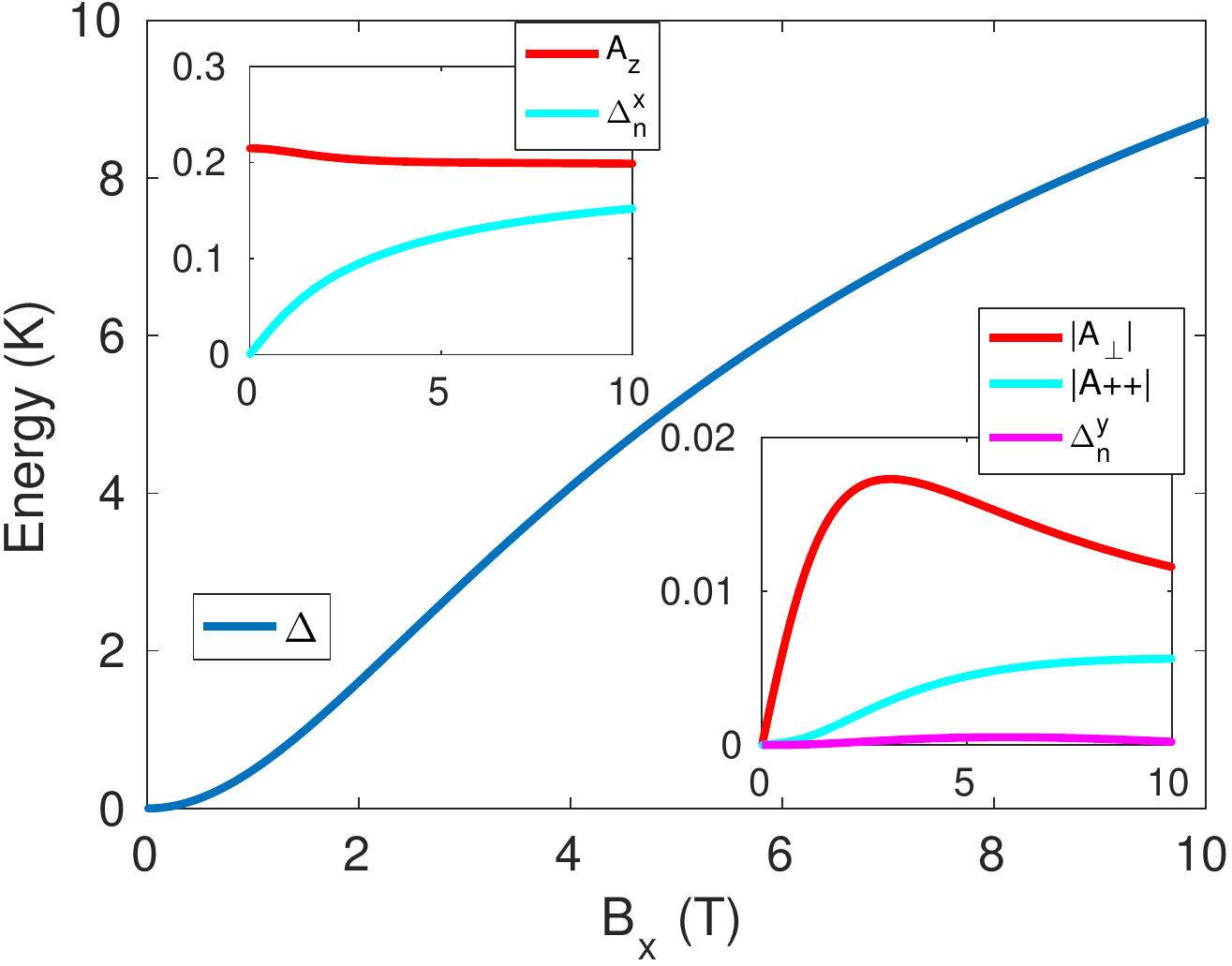}
\caption{The effective transverse field $\Delta$ (in Kelvin) mixing the Ising spins in $LiHoF_4$, as a function of the physical transverse field $B_x$ (in Tesla). The upper left inset shows the next largest parameters in the $LiHoF_4$ Hamiltonian: the longitudinal hyperfine coupling $A_z$, and the effective transverse field $\Delta_n^x$ acting directly on the nuclear spins. The lower right inset illustrates the magnitudes of the remaining transverse hyperfine parameters: $A_{\perp}$ (uppermost line), $A_{++}$ (middle line), and the stray field $\Delta_n^y$, acting on the nuclear spins in the direction transverse to the easy axis and the direction of the applied transverse field (lower line).}
\label{fig:P1}
\end{figure}

We have already seen in Fig. (\ref{fig:ME}) how the matrix elements of the effective spin operators depend on the physical transverse field $B_x$. In Fig. (\ref{fig:P1}), we show the $B_x$ dependence of the effective transverse field $\Delta$ mixing the electronic Ising spins, along with the $B_x$ dependence of the parameters in the nuclear spin Hamiltonian $\mathcal{H}_{NS}$. The longitudinal term $A_z$ dominates the hyperfine interactions, with a substantial effective transverse field $\Delta_n$ directly mixing the nuclear spins. The remaining parameters in our model are much smaller.


Combining the electronic and nuclear contributions, we find the full effective Hamiltonian to be
\begin{equation}
{\cal H}_{eff} \;=\; \mathcal{H}_e + \mathcal{H}_{NS}.
 \label{HeffT}
\end{equation}
with these two terms given by (\ref{eq:Hetrunc1}) and (\ref{eq:EffectiveHamiltonian}).

The important points to take from the low energy effective Hamiltonian $\mathcal{H}_{eff}$ are (i) the Ising nature of the system at low temperatures, (ii) the anisotropy of the truncated hyperfine interaction, and (iii) the large effective transverse magnetic field acting directly on the nuclear spins. The effective longitudinal hyperfine interaction is $A_z \sim 200mK$; the transverse component $A_{\perp}$ is over ten times smaller (this anisotropy was noted by Mennenga \textit{et al} in their specific heat measurements in 1984 \cite{Mennenga}). The effective transverse field acting on the nuclear spins $\Delta_n^x$ is roughly $100mK$ when the physical transverse field $B_x$ is between 3T and 6T. It is this effective transverse field $\Delta_n^x$, rather than the transverse hyperfine interaction $A_{\perp}$, that is mainly responsible for the mixing of the nuclear spin states.

Finally, let us note again in what way this effective Ising Hamiltonian for $LiHoF_4$ is different from the toy model. The main differences are (a) the involvement of spin-$7/2$ bath spins, instead of spin-$1/2$ two-level systems; (b) the existence of extra fields acting on these bath spins, which gives them their own dynamics, independent of that given to them by the electronic spins; and (c) the long-range dipolar interactions between the electronic spins. Perhaps needless to say, the field dependence of the various parameters in ${\cal H}_{eff}$ is now non-trivial and different from that in the toy model. 

In what follows we now wish to understand the behaviour of the phase diagram of this system, and where appropriate, compare it with that of the toy model.


\section{The Partition Function}
\label{sec:FormToy}


In this section we describe the main techniques used to derive results in this paper. To incorporate fluctuations we use a well-established technique for setting up a field-theoretical description of the system, separating off fluctuations from mean field terms in the partition function. A cumulant expansion is used to obtain the final form of the effective Hamiltonian for the fluctuations.

\subsection{Field-Theoretic form for the Partition Function}
\label{sec:partF}

The starting point for our field-theoretic formulation will be a mean field theory (MFT). This takes the usual form - we divide the Hamiltonian according to
\begin{equation}
{\cal H} \;=\; \mathcal{H}_{MF} \;+\; {\cal H}_{fl}
 \label{H-MFT}
\end{equation}
where ${\cal H}_{MF}$ is the mean field term, and we and write the fluctuation term as
\begin{equation}
\mathcal{H}_{fl}(\tau) = - \;
\frac{1}{2}\sum_{i \neq j} U_{ij}\ \delta S_i^z (\tau) \delta S_j^z (\tau),
\end{equation}
where the $S_i^z = \tau_i^z / 2$ are electronic spin operators, and $\delta S_j^z = (S_j^z - \langle S^z \rangle_0 )$ are fluctuations about the mean polarization $\langle S^z \rangle_0$ determined by $\mathcal{H}_{MF}$, leaving the MF interaction $U_0 \langle S^z \rangle_0 \sum_i S_i^z$ acting on individual sites. The fluctuation theory is developed in terms of the spin operators ${\bf S}_j$, rather than the Pauli matrices, to avoid confusion between the Pauli matrices and imaginary time indices, and to allow for easy generalization to larger spins $S > 1/2$.

It is well known that fluctuations can cause MFT/RPA results to fail near any phase transition \cite{GoldenfeldBook, SachdevBook}. Thus, if we really want to understand the QPT, we must include these, and the interactions between them. This requires an expansion of the free energy in powers of these fluctuations. To do this we write the quantum partition function in the Matsubara representation; it is then given by
\begin{equation}
Z = Z_{MF} \biggr\langle T_{\tau}\exp
\biggr[-\int_{0}^{\beta} d\tau \mathcal{H}_{fl}(\tau) \biggr] \biggr\rangle_{0},
\end{equation}
where $Z_{MF} = \text{Tr}[e^{-\beta H_{MF}}]$, and $\langle T_{\tau} \cdots \rangle_{0}$ is an imaginary time ordered thermal average taken with respect to $\mathcal{H}_{MF}$. The imaginary time dependence of the quantum operators follows from $O(\tau) = e^{\tau \mathcal{H}_{MF}} O e^{-\tau \mathcal{H}_{MF}}$. 

We decouple the interaction between the Ising spins using the Hubbard-Stratonovitch transformation \cite{HubbardTransformation, Stratonovich, MZ, YoungQPT}. Introducing an auxiliary scalar field $\phi_i(\tau)$ at each site, the partition function becomes
\begin{align}
\label{eq:PF}
Z = &Z_{MF} \int \mathcal{D}\phi
\exp \biggr(-\frac{1}{2\beta}\int_{0}^{\beta} d\tau \sum_{i} \phi_{i}^{2}\biggr)
\\ \nonumber
& \times \biggr\langle T_{\tau} \exp\biggr(-\frac{1}{\beta}\int_{0}^{\beta}
d \tau \sum_{i \neq j} \phi_{i} \sqrt{\beta U_{ij}} \delta S_{j}^z
\biggr) \biggr\rangle_{0},
\end{align}
where the intergration measure  is $\mathcal{D}\phi = \prod_i d\phi_i(\tau)  / \sqrt{2\pi}$. The imaginary time dependence of the auxiliary fields $\phi_{i}(\tau)$ and the spin operators $\delta S_{j}^z(\tau)$ are suppressed for brevity.

It is advantageous at this point to establish a relationship between the auxiliary fields $\phi_i(\tau)$ and the connected longitudinal imaginary time ordered correlation function or Green's function $G_{ij}(\tau-\tau') = -\langle T_{\tau}\ \delta S_i^z(\tau) \delta S_{j}^z(\tau')\rangle$.  We add a site and time dependent longitudinal field to the fluctuating part of the Hamiltonian
\begin{equation}
\beta \widetilde{\mathcal{H}}_{fl}(\tau) = - \; \beta \mathcal{H}_{fl}(\tau)
+ \sum_i h_i(\tau)\ \delta S_i^z(\tau),
\end{equation}
and proceed with the Hubbard-Stratonovich transformation as in (\ref{eq:PF}). Next, we shift the auxiliary fields (written here in momentum space), $\phi_{\bf k}(\tau) \rightarrow \phi_{\bf k}(\tau) - h_{\bf k}(\tau) / \sqrt{\beta U_{\bf -k}}$, to transfer the dependence of the partition function on the longitudinal field to the Gaussian prefactor \cite{WangEvensonSchrieffer}. The result is
\begin{align}
\frac{Z}{Z_{MF}} &= \int \mathcal{D} \phi
\exp\biggr(-\frac{1}{2\beta}\int_{0}^{\beta} d\tau
\sum_{\bf k} \biggr|\phi_{\bf k}-\frac{h_{\bf k}}{\sqrt{\beta U_{\bf -k}}}\biggr|^{2}\biggr)
\\ \nonumber
& \times \biggr\langle T_{\tau} \exp\biggr(-\frac{1}{\beta}\int_{0}^{\beta} d \tau
\sum_{\bf k} \phi_{\bf -k} \sqrt{\beta U_{\bf k}}\ \delta S^z_{\bf k} \biggr\rangle_{0}.
\end{align}
With the partition function written in this form we are free to take a functional derivative in order to determine the fluctuations about the MF magnetization
\begin{align}
\label{eq:MagCorr1}
\langle \delta S_{\bf k}^z(\tau)\rangle
= - \frac{\delta \ln{Z}}{\delta h_{\bf k} (\tau)} \biggr|_{h=0}
= \frac{1}{\sqrt{\beta U_{\bf k}}} \biggr\langle \phi_{\bf k}(\tau) \biggr\rangle_{\phi},
\end{align}
with the average on the left $\langle \cdots \rangle$ taken with respect to the Ising and bath spin Hamiltonian, and the average on the right $\langle \cdots \rangle_{\phi}$ taken with respect to the partition function for the auxiliary fields $Z_{\phi} = Z / Z_{MF}$. The cumulant Green's function, defined by $G_{ij}^c(\tau-\tau') = G_{ij}(\tau-\tau') + \langle\delta S_i^z(\tau) \rangle \langle \delta S_{j}^z(\tau') \rangle $, follows from
\begin{align}
G_{\bf k}^c(\tau-\tau') = - \frac{\delta \ln{Z} }
{ \delta h_{\bf -k}(\tau)\delta h_{\bf k} (\tau')} \biggr|_{h=0}
\end{align}
Performing the derivatives, and transforming to Matsubara frequency space ($\omega_r = 2\pi r / \beta$)
\begin{align}
G_{\bf k}^c(i\omega_r) =  \int_0^{\beta} d\tau e^{i\omega_r \tau}
\frac{1}{N} \sum_{ij} e^{i{\bf k} \cdot ({\bf r}_i-{\bf r}_j)} G_{ij}^c(\tau)
\end{align}
we find that
\begin{align}
\label{eq:GFc}
G_{\bf k}^c(i\omega_r) =
-\frac{1}{U_{\bf k}} \biggr[ \langle |\phi_{\bf k}(i\omega_r)|^2\rangle_{\phi} - 1 \biggr],
\end{align}
with the $\phi_{\bf k}(i\omega_r)$ defined by
\begin{align}
\phi_{\bf k}(i\omega_r) = \frac{1}{\beta} \int_0^{\beta} e^{i\omega_r \tau}
\frac{1}{\sqrt{N}} \sum_j e^{i{\bf k} \cdot {\bf r}_j}\phi_j(\tau).
\end{align}
Note that $\phi_{\bf k} (i\omega_r) = \phi_{\bf -k}^*(-i\omega_r)$ meaning the functional integral for the auxiliary field partition function double counts each degree of freedom.

We have established a general relationship between the correlations of the auxiliary field $\phi$, and the correlations of the Ising spin operators in the underlying Hamiltonian. To proceed, we must perform the thermal average $\langle \cdots \rangle_0$ in the quantum partition function (\ref{eq:PF}); the general idea is to rewrite $Z_{\phi} = Z/Z_{MF}$ as a functional integral of form
\begin{equation}
Z_{\phi} \;=\; \int \mathcal{D}\phi e^{-\beta \mathcal{H}_{eff}[\phi]}
 \label{Z-Heff}
\end{equation} 
over an effective Hamiltonian $\mathcal{H}_{eff}[\phi]$. This one does using a cumulant expansion.

\subsection{Cumulant Expansion}

To carry out a cumulant expansion of the partition function, we define the cumulants $M_n$ in the usual way \cite{Kubo}, as
\begin{align}
M_n(x) \;=\; \langle x^n \rangle \;\;- \sum_{n_1+\ldots+n_k =n}M_{n_1}M_{n_2}\ldots M_{n_k}.
\end{align}
Then, for $Z=\langle T_{\tau}\exp(\int_0^{\beta} d\tau f(\tau)) \rangle_0$, we have
\begin{align}
\ln{Z} = \sum_{n} \frac{1}{n!} \int_0^{\beta} d\tau_1 \ldots \int_0^{\beta} d\tau_n
M_n \biggr(T_{\tau} f(\tau_1) \ldots f(\tau_n)\biggr) .
\end{align}
For the auxiliary field partition function $Z_{\phi} = Z / Z_{MF}$ this becomes
\begin{align}
\label{eq:Z-Mn}
Z_{\phi} &= \int \mathcal{D}\phi
\exp\biggr(-\frac{1}{2\beta}\int_{0}^{\beta} d\tau \sum_{\bf k} |\phi_{\bf k}(\tau)|^{2}\biggr)
\\ \nonumber
& \times T_\tau \biggr[
\exp\biggr(\sum_{n=1}^{\infty}\frac{(-1)^n}{n!\beta^n} \prod_{i=1}^n \int_{0}^{\beta} d \tau_i
\bigr\langle M_n\bigr(V(\tau)\bigr)\bigr\rangle_{0}\biggr)\biggr] ,
\end{align}
where $M_n$ is now the n$^{th}$ cumulant of $V(\tau) = \beta \mathcal{H}_{fl}(\tau)$. In calculating the cumulants we extract the auxiliary fields from the $M_n$ to write everything in terms cumulants of the spin operators; thus eg., for $M_2$ we have
\begin{align}
\langle M_2(V(\tau)) \rangle_0 =
T_{\tau}\biggr[\sum_{\bf k,k'} &\phi_{\bf -k}(\tau_1)\phi_{\bf -k'}(\tau_2) \beta \sqrt{U_{\bf k} U_{\bf k'}}
\\ \nonumber
&\times \langle M_2(\delta S_{\bf k}^z(\tau_1)\delta S_{\bf k'}^z(\tau_2))\rangle_0 \biggr].
\end{align}
When an average is performed over a cumulant of a set of statistically independent operators (for example, the $S_j^z$ at different sites are independent with respect to the probability distribution determined by the MF Hamiltonian) the cumulant of products of the Fourier transformed operators $S_{\bf k}^z =\frac{1}{\sqrt{N}}\sum_j e^{i{\bf k} \cdot {\bf r}_j} S_j^z$ reduce as follows:
\begin{align}
\label{eq:reduction}
\biggr\langle M_n(&\delta S_{\bf k_1}^z(\tau_1)\delta S_{\bf k_2}^z(\tau_2)
\ldots \delta S_{\bf k_n}^z(\tau_n)) \biggr\rangle_0 =
\\ \nonumber
&\frac{1}{N^{\frac{n-2}{2}}}
\biggr\langle
M_n(S^z(\tau_1)S^z(\tau_2) \ldots S^z(\tau_n)) \biggr\rangle_0
\delta_{\sum_{i=1}^n {\bf k}_i,0}
\end{align}
The cumulant on the right hand side contains spins at a single site at multiple imaginary time indices. This reduction leads to a significant simplification in the effective Hamiltonian $\mathcal{H}_{eff}[\phi]$, which we can now write as
\begin{align}
\label{eq:HQ-eff}
\mathcal{H}_{eff}[\phi] = \frac{1}{\beta}
\sum_{n=1}^{\infty}\biggr[ \sum_{\{ r_i, {\bf k}_i \}}
\frac{u_n (\{ {\bf k}_i \},\{ i\omega_{r_i} \})}{n!}
\prod_{i=1}^n \phi_{{\bf k}_i}(i\omega_{r_i}) \biggr],
\end{align}
where the functions $u_n (\{ {\bf k}_i \},\{ i\omega_{r_i} \})$ are coupling constants in the effective Hamiltonian between the fields $\phi_{\bf k}(i\omega_r)$.

We can calculate expressions for these interaction functions: making use of the reduction (\ref{eq:reduction}), and comparing (\ref{eq:HQ-eff}) to (\ref{eq:Z-Mn}), we have
\begin{align}
u_2 = \biggr[ \delta_{i\omega_{r_1},-i\omega_{r_2}}
-\frac{\sqrt{U_{{\bf k}_1}U_{{\bf k}_2}}}{\beta}  M_2(-i\omega_{r_1},-i\omega_{r_2})
\biggr] \delta_{{\bf k}_1,-{\bf k}_2},
 \label{eq:u2}
\end{align}
for the term quadratic in the auxiliary fields, and
\begin{align}
\label{eq:un}
u_n =  \frac{(-1)^{n+1}}{N^{\frac{n}{2}-1}}
\biggr[\prod_{i=1}^n (\beta U_{-{\bf k}_i})^{\frac{1}{2}}\biggr]
\frac{1}{\beta^n} M_n(\{-i\omega_{r_i}\}) \delta_{\sum {\bf k}_i , 0}
\end{align}
for the higher order terms. The couplings $\{ u_n \}$ between $n$-tuplets of fluctuation fields contain energy and momentum conserving $\delta$-functions

These expressions are exact; no approximations have been used to derive them. The Matsubara frequency dependence of the spin cumulants $M_n$ comes from the Fourier transform of the spin operators, which are transformed according to $S_{\bf k}^z(i\omega_r) = \int_0^{\beta} d\tau  e^{-i\omega_r \tau} S_{\bf k}^z (\tau)$. This Fourier transform convention eliminates factors of $\beta$ from the expressions for the cumulants.

In what follows, it will only be necessary to work up to quartic order in the fields, Thus, the effective Hamiltonian we will use has the form
\begin{align}
\label{eq:Heff}
\beta\mathcal{H}_{eff}[\phi] \;\;=\;\; \frac{1}{2}&\sum_{r,{\bf k}}
\; (\mathcal{D}^o_{\bf k}(i\omega_r))^{-1} \; |\phi_{\bf k}(i\omega_r)|^2
\\ \nonumber
&+ \frac{1}{3!}
\biggr[ \sum_{\{ r_i, {\bf k}_i \}}  u_3 \prod_{i=1}^3 \phi_{{\bf k}_i}(i\omega_{r_i}) \biggr]
\\ \nonumber
&\qquad +\frac{1}{4!}\biggr[ \sum_{\{ r_i, {\bf k}_i \}}
u_4 \prod_{i=1}^4 \phi_{{\bf k}_i}(i\omega_{r_i})\biggr],
\end{align}
where we have suppressed the momentum and frequency dependence of the $u_n \equiv u_n(\{ {\bf k}_i \},\{ i\omega_{r_i} \})$. 

The first term here is of course just the Gaussian, or random phase approximation (RPA) \cite{GoldenfeldBook}, with the free field propagator $\mathcal{D}^o_{\bf k}(i\omega_r)$ approximation to the full propagator, in which we allow fluctuations about the MF, but treat these fluctuations as non-interacting. The second order spin cumulant is $\sum_{r_2} M_2(-i\omega_{r_1},-i\omega_{r_2}) = \beta g(i\omega_r)$, where $g(\tau) = \langle \delta S^z(\tau) \delta S^z(0) \rangle_0$ is the MF Green's function (see Appendix \ref{ap:Cumulants}). One has 
\begin{align}
\label{eq:FFP}
\mathcal{D}^o_{\bf k}(i\omega_r) = \frac{1}{1+g(i\omega_r)U_{\bf k}}.
\end{align}
and we have $\langle |\phi_{\bf k}(i\omega_r)| \rangle_{\phi} = \mathcal{D}^o_{\bf k}(i\omega_r)$. At low energies and small momenta the free field propagator may be expanded as $(\mathcal{D}_{\bf k}^o(i\omega_r))^{-1} = r_0 + \alpha_1 {\bf k}^2 + \alpha_2 k_z^2/ {\bf k}^2 + \gamma (i\omega_r)^2 + \cdots$, where $r_0$, $\alpha_i$, and $\gamma$ depend on the nature of the interaction between spins and other details of the model in question.

To complete the theory, and to determine the effective Hamiltonian $\mathcal{H}[\phi]$, we must also calculate the cumulants $M_n(\{i\omega_{r_i}\})$. These are now cumulants of spins at a single site (at multiple imaginary time indices) determined with respect to $\mathcal{H}_{MF}$. These calculations are straightforward; however, they are rather lengthy, as are the final expressions. The results are therefore given in Appendix \ref{ap:Cumulants}.


\section{RPA Susceptibilities, Correlation Functions, and Eigenmodes}
\label{sec:CorrFuncs}


Before using the full 4th-order expansion in (\ref{eq:Heff}), we first derive the results in the RPA; as just discussed, this is defined by the Gaussian approximation, ie., the first term in (\ref{eq:Heff}). Thus, in this section we will derive results for the RPA dynamic susceptibility. This will be done in the first two sub-sections for the Ising spins, first for the Toy Model, and then for $LiHoF_4$ (where the Ising susceptibility is just the electronic spin susceptibility). Finally, in the last sub-section, we calculate the ``total susceptibility", which includes contributions from the bath spins as well.

\subsection{Results for Ising spins}

We will be interested in this sub-section in the susceptibility if the Ising spins, ie., we are interested in
\begin{equation}
G_{\bf k}(\tau) = -\langle \delta \tau_{\bf k}^z(\tau)\ \delta \tau_{\bf -k}^z (0) \rangle
\end{equation}

From (\ref{eq:GFc}), and the fact that at the RPA level of approximation $G_{\bf k}^c(i\omega_r) = G_{\bf k}(i\omega_r)$, we see that one has
\begin{align}
\label{eq:Gk}
G_{\bf k}(i\omega_r) = \frac{g(i\omega_r)}{1+V_{\bf k} g(i\omega_r)}
\end{align}
with $g(i\omega_r)$ being the Fourier transform of the MF Green's function $g(\tau) = \langle \delta \tau^z (\tau) \delta \tau^z (0)\rangle_0$.

We wish to give results for (\ref{eq:Gk}) in terms of parameters in $\mathcal{H}_{MF}$ for both (i) the toy model, and (ii) the $LiHoF_4$ system. The dynamic susceptibility of the Ising spins follows from the analytic continuation $\chi_{\bf k}^{zz}(\omega) = -G_{\bf k}(i\omega_r \rightarrow \omega + i\epsilon)$, with $\epsilon$ being a small constant later taken to zero. Thus we have direct access to the RPA eigenmodes (the poles of the Green's function), which correspond to the low energy states of the system, and their corresponding spectral weights.

\subsubsection{The Toy Model}
 \label{corrToy}
 
In the case of the toy model, with a spin half Ising lattice coupled to a set of spin half bath spins, we will start from a Hamiltonian
\begin{align}
{\cal H}  =  -\sum_{i<j} &V_{ij}\tau_i^z \tau_j^z
-h \sum_j \tau_j^z - \Delta_{0} \sum_j \tau_j^x
\\ \nonumber
&\;+\; A_{z} \sum_j  \sigma_j^z \tau_j^z
+ \frac{A_{\perp}}{2} \sum_{j} (\sigma_j^+ \tau_j^- + \sigma_j^- \tau_j^+),
\end{align}
which differs in form from the Hamiltonian previously derived for the toy model (ie., that given in (\ref{Htoy})) only by the addition of a longitudinal field term $h \sum_j \tau_j^z$. Let us now rewrite this as
\begin{equation}
\label{eq:H-MFT}
{\cal H} \;=\; {\cal H}_{MF} \;-\; \sum_{i < j}
V_{ij}\ \delta \tau_i^z  \delta \tau_j^z,
\end{equation}
where $\delta \tau_j^z = \tau_j^z - \langle \tau_j^z \rangle_0$, and where the MF Hamiltonian ${\cal H}_{MF} = \sum_j {\cal H}_{MF}^j$, with single-site terms of form
\begin{align}
{\cal H}^j_{MF} \;=\; - \Delta_{0}\tau_j^x &-
\bigr(h + V_0 \langle \tau^z \rangle_0 \bigr)  \tau_j^z
\\ \nonumber
&\;+\;  A_{z} \sigma_{j}^{z} \tau_{j}^{z}
+ \frac{A_{\perp}}{2} (\sigma_{j}^{+} \tau_{j}^{-} + \sigma_{j}^{-} \tau_{j}^{+}).
\end{align}
The RPA calculation of Green's function then consists in treating the fluctuation term in (\ref{eq:H-MFT}) in a Gaussian approximation.

At this point it is useful to introduce the eigenstates $|n\rangle$ of the MF Hamiltonian, so that
\begin{equation}
{\cal H}_{MF}^j|n\rangle  \;=\; E_n |n\rangle
 \label{Hmf-n}
\end{equation}
and also define the quantities
\begin{equation}
c_{mn} = \langle m| \tau^z|n \rangle_0
 \label{c-mn}
\end{equation} 
as the MF matrix elements of the Ising spin operator.

The result for the RPA Green's function then follows from that for the MF Green's function $g(\tau) = \langle \tau^z(\tau) \tau^z(0) \rangle_0$, which we derive making use of the Hubbard operator formalism \cite{Hubbard, Yang} in Appendix \ref{ap:Cumulants}. The result at Matsubara frequency $\omega_r$ and inverse temperature $\beta$ is
\begin{widetext}
\begin{align}
\label{eq:g}
g(i\omega_{r}) = - \sum_{n > m} c_{mn}^2 p_{mn}
\frac{2 E_{nm}}{E_{nm}^2-(i\omega_{r})^2}
-  \beta \biggr(\sum_{m}c_{mm}^2p_m - \biggr[\sum_m c_{mm}p_m \biggr]^2\biggr)     \delta_{\omega_r,0},
\end{align}
\end{widetext}
where $E_{nm} = E_n-E_m$ is the difference between energy levels of $\mathcal{H}_{MF}$, and the $p_{mn} = p_m - p_n$ with $p_m = e^{-\beta E_m}/\sum_n e^{-\beta E_n}$ are population factors.  The second (elastic) contribution to (\ref{eq:g}) vanishes in the paramagnetic phase, and in the limit $T \rightarrow 0$. The MF susceptibility of the system is given by $\chi_0^{zz} = -g(0)$. The longitudinal RPA Green's function $G({\bf k},z) = g(z)/(1 + V_{\bf k} g(z))$, evaluated at $T=0$, is then
\begin{widetext}
\begin{align}
\label{eq:GFRPA}
G({\bf k},z)\biggr|_{T=0} =
\frac{\sum_{n=2}^{4}c_{1n}^2 2E_{n1} \prod_{m\neq n} (E_{m1}^{2}-z^2)}
{V_{\bf k}\sum_{n=2}^{4}c_{1n}^2 2E_{n1} \prod_{m\neq n} (E_{m1}^{2}-z^2)
- \prod_{n=2}^{4} (E_{n1}^{2}-z^2)}.
\end{align}
\end{widetext}
The RPA modes of the system $\{E_{\bf k}^p\}$ follow from the poles of this function. Writing the dynamic susceptibility $\chi_{\bf k}^{zz}(\omega) = -G_{\bf k}(i\omega_r \rightarrow \omega + i\epsilon)$ as $\chi_{\bf k}^{zz}(\omega) = \chi_{\bf k}'(\omega)+i \chi_{\bf k}''(\omega)$, the spectral weight of the $p^{th}$ RPA mode $A_{\bf k}^p$ follows from $\chi_{\bf k}''(\omega) = \sum_p A_{\bf k}^p \delta(\omega - E_{\bf k}^p)$. 

\begin{figure}[htp]
\centering
\includegraphics[width=8.5cm]{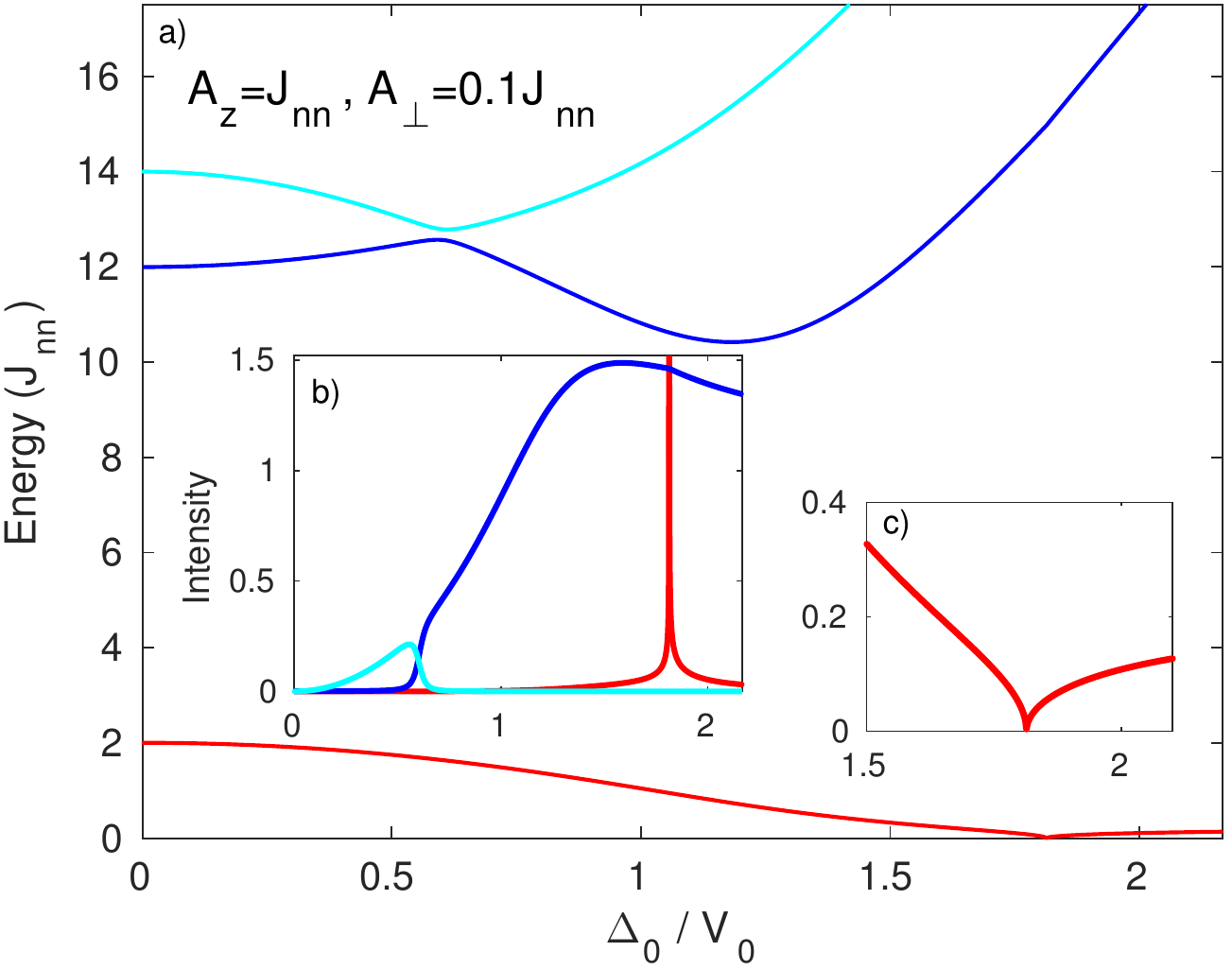}
\caption{(a) Energies (in units of the nearest neighbour exchange coupling $J_{nn}$) of the zero temperature RPA modes $\{E_p \}$ of the toy model with an anistropic hyperfine interaction, at ${\bf k}=0$, in 3-dimensions ($V_0=6J_{nn}$), as a function of the normalized transverse field $\Delta_0 / V_0$. (b) Intensities of the RPA modes $A_p^z$ (arbitrary units), with $\chi_{zz}''(0,\omega) = \sum_p A_p^z \delta(\omega-E_p)$, associated with the longitudinal Ising spin susceptibility. The RPA modes are colour coordinated with their intensities. (c) Energy of the soft mode in the vicinity of the quantum critical point.}
\label{fig:ElectronicModes1}
\end{figure}

These results are used to produce Fig (\ref{fig:ElectronicModes1}), in which we show these eigenmodes and associated spectral weights for a strongly anisotropic hyperfine coupling with a dominant longitudinal component, at zero temperature. The key features here are (i) a low-energy collective mode which softens to zero energy at the QPT, with sharply-peaked spectral weight near the QPT, and (ii) a clear effect of the QPT on the higher modes as well.

The zero temperature critical transverse field is determined by the point at which the RPA susceptibility diverges at zero wavevector and frequency, that is, when $1-V_0 \chi_0^{zz} = 0$. Above $\Delta_c$, the MF susceptibility may be written
\begin{align}
\chi_0^{zz}\biggr|_{T=0} =
\frac{2c_{12}^2}{E_{21}}+\frac{2c_{14}^2}{E_{41}}
\qquad \Delta_0 > \Delta_c.
\end{align}
In the high field limit $\Delta_0 \gg A_z, A_{\perp}, V_0$, we may expand to $O(A_{z, \perp}^3 / \Delta_0^3, V_0^3 / \Delta_0^3)$ to obtain
\begin{align}
\Delta_c = \frac{V_0}{4}+\frac{V_0}{4}\sqrt{1+\frac{4}{V_0}
\biggr(\frac{A_z^2}{A_{\perp}}-A_{\perp}\biggr)}.
\end{align}
We see that as the hyperfine interaction becomes increasingly anisotropic, with $A_z > A_{\perp}$, the critical transverse field becomes increasingly large.


Let us recall again that these calculations are done at temperature $T=0$. Although we will not do finite $T$ calculations in this paper, it is worthwhile noting that if we carry out the same calculation at finite $T$, we will find six eigenmodes, with the additional three corresponding to transitions between excited states; again, the QPT shows up in the low energy hybridized mode between the Ising and bath spins, with sharply peaked spectral weight when $\Delta \sim \Delta_c$.

Let us also emphasize again that these MFT/RPA results for the toy model do not yet account for ``mode-mode interactions" between fluctuations about the MF eigenstates - we do this in the next section.

\subsubsection{The LiHoF$_4$ System}
 \label{sec:corrLiHo}

The MF part of the full truncated Hamiltonian for the $LiHoF_4$ system at a single-site is given by
\begin{align}
\mathcal{H}_{MF}^i = - \frac{\Delta}{2}  &\tau_{i}^{x}
- V_0 \langle \tau^{z} \rangle_0  \tau_{i}^{z}
+ \vec{\Delta_{n}} \cdot \vec{I_{i}} + A_{z} \tau_{i}^{z}I_{i}^{z}
\\ \nonumber
&+\left( A_{\perp}  \tau_{i}^{+}I_{i}^{-}
+ A_{++}  \tau_{i}^{+}I_{i}^{+} + h.c. \right)
\end{align}
in which $V_0 =  C_{zz}^2 [ J_{D}  \sum_j D_{ij}^{zz} - 4 J_{nn} ]$ is the ${\bf k}=0$ limit of $V_{\bf k}$, and where all the other parameters were discussed in Section \ref{sec:LiHo} above. Again we define eigenstates $|n\rangle$ and eigenenergies $E_n$ for this MF Hamiltonian, but now in an enlarged 16-dimensional Hilbert space incorporating the 2 electronic states and the 8 nuclear states for each site.

The development of the RPA forms for the correlators then proceeds as for the toy model. We get a longitudinal RPA spin correlation function
\begin{align}
G({\bf k},z) 
= \langle \delta J^z(z) \delta J^z(-z)\rangle_0= \frac{C_{zz}^2 g(z)}{1+ V_{\bf k}  g(z)},
\end{align}
with $g(\tau) = -\langle T_{\tau} \delta \tau^z (\tau) \delta \tau^z (0) \rangle_0$. The RPA Green's function then becomes
\begin{widetext}
\begin{align}
\label{eq:GFRPA2}
G({\bf k},z)\biggr|_{T=0} =
\frac{ C_{zz}^2  \; \sum_{n=2}^{16}c_{1n}^2 2E_{n1} \prod_{m\neq n} (E_{m1}^{2}-z^2)}
{V_{\bf k}\sum_{n=2}^{16}c_{1n}^2 2E_{n1} \prod_{m\neq n} (E_{m1}^{2}-z^2)
- \prod_{n=2}^{16} (E_{n1}^{2}-z^2)}.
\end{align}
\end{widetext}
ie., it has the same form as in (\ref{eq:GFRPA}) except the summations and products are now over 16 MF energy levels (with the matrix element $c_{mn} = \langle m| \tau^z|n \rangle_0$ now defined between these states); and there is an additional prefactor $C_{zz}^2$ coming from the truncation procedure.

\subsection{Total Dynamic Susceptibility}
\label{sec:TotalSusc}

In an Ising system coupled to a spin bath the dynamic susceptibility will contain contributions from both the Ising spins and the bath spins. Here we give RPA results for the total susceptibility of Ising plus bath spin systems. The total dynamic susceptibility then has the form
\begin{align}
\chi_{{\bf k}}^{\mu\nu}(t) = i\theta(t)
\biggr\langle \biggr[ \delta \tau_{\bf k}^{\mu} (t)+\gamma \delta \sigma_{\bf k}^{\nu} (t),
\ \delta \tau_{\bf k}^{\mu} (0)+\gamma \delta \sigma_{\bf k}^{\nu} (0) \biggr] \biggr\rangle,
\end{align}
where $\mu$ and $\nu$ may equal $x,y$, or $z$, and $\gamma$ is the ratio of bath spin and Ising spin gyromagnetic ratios. In $LiHoF_4$ we have $\gamma = g_n \mu_n / g_L \mu_B \approx 1/550$, and even at energies corresponding to the hyperfine splitting the response is dominated by the electronic contribution. However one can imagine more general scenarios in which $\gamma$ is much larger, and so it is useful to derive the results which follow.

The dynamic response functions follow from the imaginary time correlation functions $\chi_{{\bf k}}^{\mu\nu}(\omega) = -G^{\mu\nu}({\bf k}, i\omega \rightarrow \omega + i \epsilon)$, which we write as follows:
\begin{align}
G^{\mu\nu}({\bf k}, &z) \;\; = \;\; G_{\tau\tau}^{\mu\nu}({\bf k}, z)
\\ \nonumber
&+ \gamma \biggr(G_{\sigma\tau}^{\mu\nu}({\bf k}, z)+G_{\tau\sigma}^{\mu\nu}({\bf k}, z)\biggr)
+\gamma^2  G_{\sigma\sigma}^{\mu\nu}({\bf k}, z).
\end{align}
In the RPA, the correlation functions are given by \cite{Stinchcombe1}
\begin{align}
\label{eq:corr}
G_{ab}^{\mu\nu} ({\bf k},z)\biggr|_{RPA} \;= \; g_{ab}^{\mu \nu} (z)
- \frac{g_{a\tau}^{\mu z}(z) V_{\bf k} g_{\tau b}^{z \nu}(z)}
{1+g_{\tau\tau}^{zz}(z)V_{\bf k}},
\end{align}
where $a$ and $b$ may refer to either a bath spin $\sigma$, or an Ising spin $\tau$. The $g_{ab}^{\mu \nu} (z)$ are connected MF imaginary time correlation functions; for example,
\begin{align}
g_{\sigma \tau}^{\mu \nu}(\tau)
= -\bigr\langle T_{\tau} \delta \sigma^{\mu} (\tau) \delta \tau^{\nu} (0) \bigr\rangle_0,
\end{align}
with the time ordered thermal average $\langle \cdots \rangle_0$ taken with respect to the single site MF Hamiltonian $\mathcal{H}_{MF}^i$.

The RPA spectrum of a system follows from the zeros of $1+g_{\tau\tau}^{zz}(z)V_{\bf k}$; the spectral weight carried by these modes depends on the particular response function. Quite generally, we may write the spectral weight carried by the $p^{th}$ RPA mode $E_{\bf k}^p$ as
\begin{widetext}
\begin{align}
A_{ab}^{\mu \nu} ({\bf k}; p)
= \frac{\prod_{m>1}[E_{m1}^2-(E_{\bf k}^p)^2]
[g_{ab}^{\mu\nu}(E_{\bf k}^p)+g_{ab}^{\mu\nu}(E_{\bf k}^p)V_{\bf k} 
g_{\tau\tau}^{zz}(E_{\bf k}^p)
-g_{a\tau}^{\mu z}(E_{\bf k}^p)V_{\bf k} g_{\tau b}^{z\nu}(E_{\bf k}^p)]}
{2E_{\bf k}^p \prod_{s\neq p} [(E_{\bf k}^p)^2-(E_{\bf k}^s)^2]},
\end{align}
\end{widetext}
where $A_{ab}^{\mu\nu} ({\bf k}; p)$ is the residue of the $p^{th}$ pole of $G_{ab}^{\mu\nu} ({\bf k},z)$. 

\begin{figure}[htp]
\centering
\includegraphics[width=8.4cm]{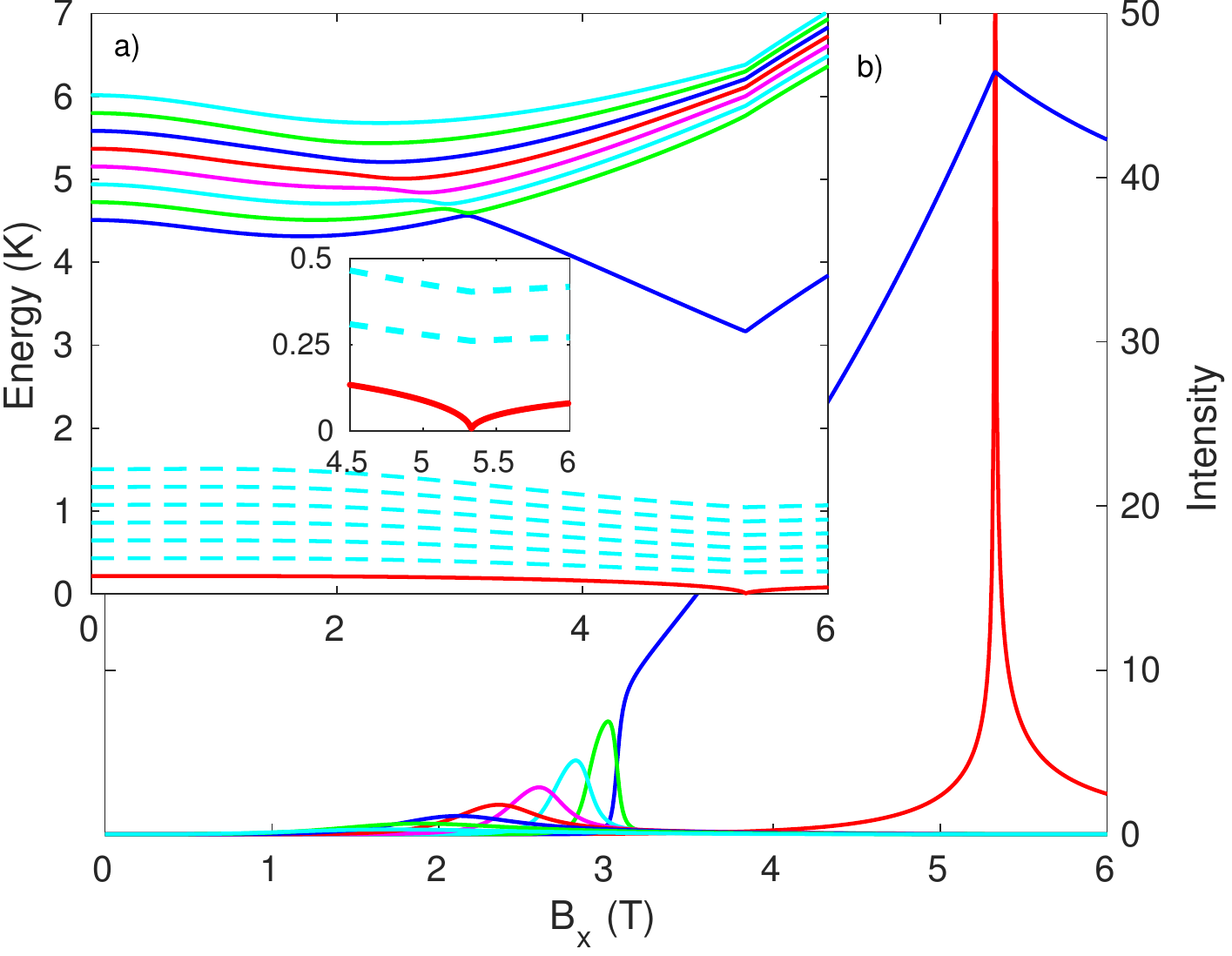}
\caption{(a) Energies (in Kelvin) of the zero temperature RPA modes of $LiHoF_4$ at ${\bf k}=0$, as a function of transverse field $B_x$ (in Tesla). The inset shows the electronuclear soft mode in the vicinity of the quantum critical point. (b) Dominant spectral weights $A_p^z$, with $\chi_{zz}''(0,\omega) = \sum_p A_p^z \delta(\omega-E_p)$, associated with the longitudinal Ising spin susceptibility (arbitrary units). The modes shown by dashed lines carry negligible spectral weight.}
\label{fig:RPASpectrumLiHo}
\end{figure}

If we now calculate MFT/RPA results for the total dynamic susceptibility of $LiHoF_4$, we find the results shown in Fig. \ref{fig:RPASpectrumLiHo}. This figure shows the low-energy collective modes of the system, and their spectral weights, as a function of transverse field $B_x$ (the dependence on the angle of the transverse field ${\bf B}_{\perp}$ in the plane is weak). The energy of the lowest electronuclear mode vanishes at the QPT, at the point where the MF magnetization vanishes.


\section{Quantum Fluctuations and the Phase Diagram}
\label{sec:Qfluc}


We now proceed to look at one of the most interesting questions in this field, viz., the effect of the bath spins on the phase diagram of the system. As briefly noted in Section \ref{sec:LiHo}, this question has been controversial; some early experiments \cite{RonnowScience05} indicated that the Quantum Ising QPT was suppressed in the $LiHoF_4$ system by the hyperfine coupling to the $Ho$ nuclear spins. Although the theory of this system clearly shows the role of electronuclear modes, it was not until very recently that these were seen experimentally \cite{KovacevicRonnow16} using NMR. In this section we intend to clear this question up theoretically; in the next section we look at the comparison with experiment.

One can get a first look at this question by looking at a mean field theory result; this we do immediately below. However it is well known that in the vicinity of any phase transition one must go beyond any RPA to get correct results - it is at this point that we must go to 4th order in a fluctuation expansion, using the field theory developed above. These results are given in the 2nd part of this section.

\subsection{Mean Field Phase Diagram}

\begin{figure}[htp]
\centering
\includegraphics[width=8.5cm]{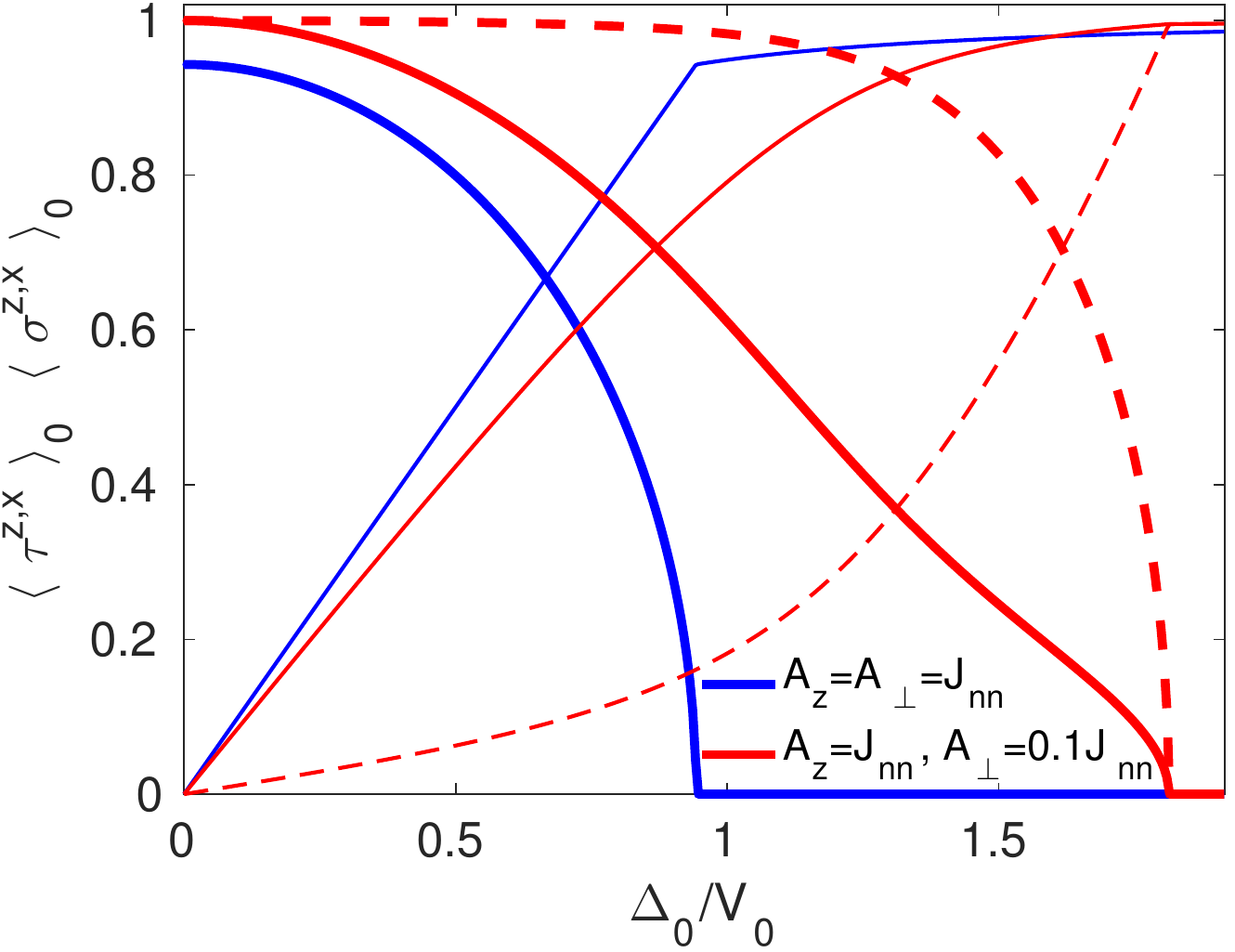}
\caption{Ground state MF polarizations of the Ising and bath spins for the toy model in 3-dimensions ($V_0=6J_{nn}$, with $J_{nn}$ being the nearest neighbour exchange coupling), as a function of the normalized transverse field $\Delta_0 / V_0$. The thick lines denote longitudinal polarizations, and the thin lines denote transverse polarizations. In blue, we see the polarizations for a system with an isotropic bath (Ising and bath spin polarizations coincide). In red, we see the electronic (solid line) and nuclear (dashed line) polarizations for the toy model with an anisotropic bath.}
\label{fig:Polarization}
\end{figure}

The MF phase diagram follows from a self consistent calculation of the longitudinal MF magnetization. Results for the $LiHoF_4$ system are very much the same as for the toy model. As an illustration, Fig. (\ref{fig:Polarization}) shows the MF results for the polarizations of both Ising and bath spins, at temperature $T=0$, for the case of the toy model where the Ising system is an  exchange coupled ferromagnet on a simple cubic lattice ($V_0=6J_{nn}$). We note the following features:

(i) There is clearly a QPT at the the critical transverse field $\Delta_c = V_0 g_c$, even when there is a spin bath. 

(ii) Any anisotropy in the hyperfine couplings has a marked effect; we see that $\Delta_c$ increases rapidly with $A_z/A_{\perp}$ (becoming infinite when $A_{\perp} \rightarrow 0$). 

This latter result can be explained by a spin bath ``blocking" mechanism \cite{SchechterStampPRL, SchechterStampPRB}. If $A_{\perp} = 0$, then at at $T=0$, with no mechanism for flipping the bath spins, the transverse field at any site $i$ is not able to mediate transitions between the degenerate states $|\Uparrow \downarrow \rangle_i$ and $|\Downarrow \uparrow \rangle_i$. The ordered bath spins then act as a longitudinal field $A_z \langle \sigma^z \rangle_0 \sum_i \tau_i^z$, which destroys the QPT. Switching on $A_{\perp}$ restores the flipping mechanism, as does going to finite temperatures, where thermal bath spin fluctuations restore the phase transition.

To summarize: we see that in mean field theory, the bath spins do not destroy the QPT, although hyperfine anisotropy profoundly affects the shape of the phase diagram. The next step is then to see how fluctuations may change these results.


\subsection{Fluctuation effects on the Phase Diagram}
\label{sec:FlucEffects}

In this sub-section we will recalculate the phase diagram, now incorporating fluctuation effects up to 4th order in the fields (cf. eqtn. (\ref{eq:Heff})). To do this we will make essential use of the results derived in the appendices. We can summarize these results as follows:

(i) We can calculate explicit expressions for the cumulants which enter into our results (\ref{eq:u2}), (\ref{eq:un}) for the interaction coefficients $u_n$ appearing in the effective free energy (\ref{eq:HQ-eff}), or its truncation as far as quartic terms in (\ref{eq:Heff}). These cumulants, defined as
\begin{align}
M_n(S_1^z \ldots &S_n^z) = \langle S_1^z \ldots S_n^z \rangle_0
\\ \nonumber
&- \sum_{n_1+\ldots+n_k =n}M_{n_1}M_{n_2} \ldots M_{n_k},
\end{align}
then have the rather complicated forms given in eqtns. (\ref{eq:M2}), (\ref{eq:M3}), and (\ref{eq:M4}) of Appendix \ref{ap:Cumulants}. 

(ii) We can then calculate an expression for the Ising spin magnetization $\langle S_z \rangle$, again at $T=0$, in terms of the cumulants. This is done perturbatively, as an expansion in powers of $1/z_c$, where $z_c$ is the coordination number of the lattice involved. The key result is that the leading correction to the MF results is from the 3rd-order cumulant, and is given by
\begin{align}
\langle S^z \rangle_1 = -\frac{\mathcal{D}_{{\bf k}=0}^o(0)}{2N \beta^2} \sum_{r,{\bf k}}  {\cal T}_{\bf k}(i\omega_{r})
\;  M_3(0,i\omega_{r},-i\omega_{r}),
\end{align}
where $\mathcal{D}_{{\bf k}=0}^o(0)$ is the zero frequency and wavevector component of the free field propagator. We have written the explicit expression for $M_3(i\omega_{r_1},i\omega_{r_2},i
\omega_{r_3})$ in terms of Bose-Matsubara frequencies $\omega_{r_j} = 2 \pi r_j / \beta$ in eqtn. (\ref{eq:M3}) of Appendix \ref{ap:Cumulants}; and ${\cal T}_{\bf k}(i\omega_{r}) = V_{\bf k} {\cal D}_{\bf k}^o(i\omega_r)$ is the renormalized RPA interaction between the Ising spins.


\begin{figure}[htp]
\centering
\includegraphics[width=8.4cm]{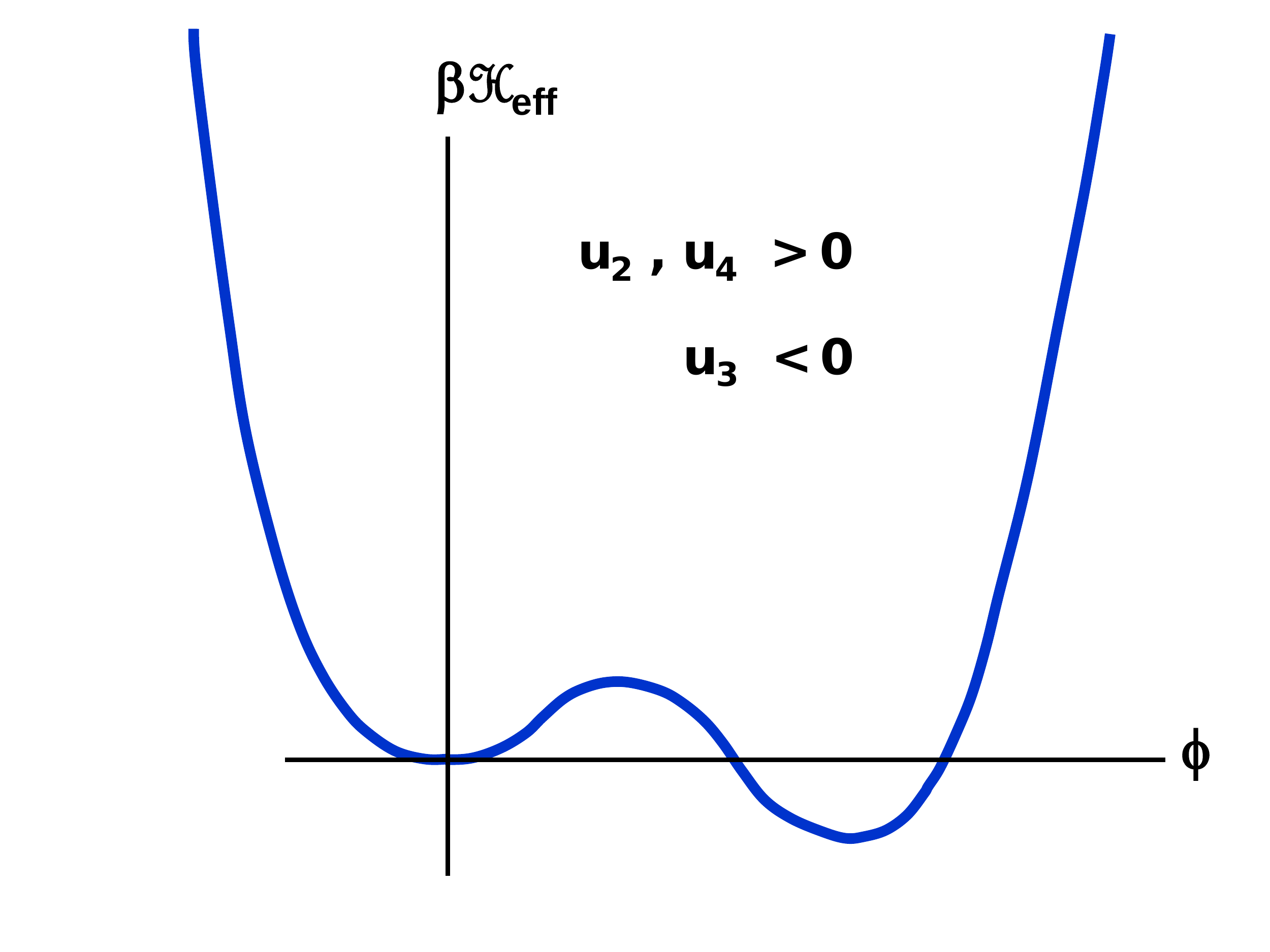}
\caption{Schematic plot of the quantity $\beta {\cal H}_{eff}(\phi)$; this can be taken as the value of the functional $\beta {\cal H}[\phi]$ of the field $\phi$, taken for some set of fixed values of the arguments of $\phi({\bf k}, i\omega_r)$. The coefficients $u_2, u_4 > 0$, in line with the calculated results, and we assume that $u_3 < 0$. The sign of $u_3$ depends on the broken $Z_2$ symmetry of the underlying Hamiltonian.}
\label{fig:Heff}
\end{figure}


In the perturbative expansion, each free momentum summation in the resulting perturbation series leads to a factor of $z_c^{-1}$. This ``high density" approximation was originally used by Brout to study random ferromagnetic systems \cite{Brout59}. For the spin half quantum Ising model, with no spin bath, the results are equivalent to those of Stinchcombe \cite{Stinchcombe1, Stinchcombe2, Stinchcombe3} (derived here in a new way), and the $1/z_c$ expansion is explained in this work. In the ordered phase, explicit calculation of $M_3$ gives corrections to the MF phase diagram of order $1/z_c$, determined by (\ref{eq:MagCorr1}). The mode-mode coupling coming from $u_4$ gives small corrections to this, as they are of order $1/z_c^3$. Explicit expressions for the leading order magnetization corrections in the quantum (T=0) regime are given in terms of parameters in $\mathcal{H}_{MF}$ in Appendix \ref{ap:MagCorrections}.

The basic structure of the results can be understood with reference to Fig. \ref{fig:Heff}. In this figure, all reference to the energy and momentum dependence of the field $\phi({\bf k},i\omega_r)$ and of the coefficients $u_n(\{ {\bf k}_i, i\omega_{r_i} \})$ is omitted; and the quantity $\beta {\cal H}_{eff}$ is then shown assuming a 4th-order truncation, as in eqtn. (\ref{eq:Heff}). Two key points emerge from the calculations:

(a) In the quantum regime (T=0), the calculations of $M_2, M_3$, and $M_4$ show that both the $u_2$ and the $u_4$ terms are always positive, ie., repulsive. Thus if $u_3 = 0$, we simply have an effective potential which increases for large fluctuations. Quartic fluctuations do not then destabilize the QPT; in fact they do the opposite. 

(b) The 3rd-order fluctuation coefficient $u_3$ is only non-zero in the ordered phase. This is in accordance with the $Z_2$ symmetry of the microscopic models. It means that in this phase, two minima are developed, the lowest of which is at a finite value of the field. 

\begin{figure}[htp]
\centering
\includegraphics[width=8.5cm]{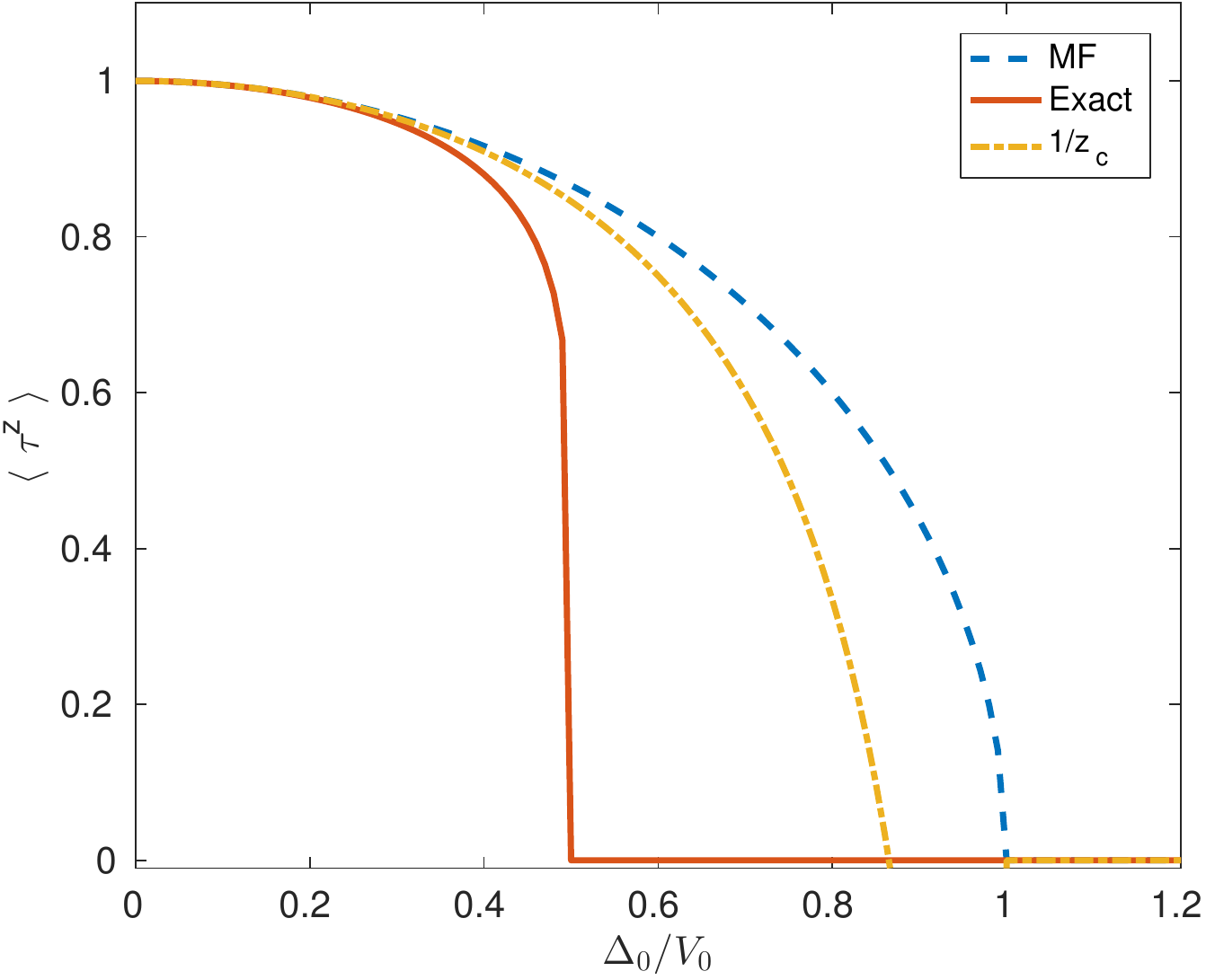}
\caption{Longitudinal spin polarization of the transverse field Ising chain ($V_0 = 2J_{nn}$). We compare the exact solution with the solution obtained in MF theory, and the leading order correction to MF theory obtained in the high density approximation.}
\label{fig:TFIM1D}
\end{figure}

The accuracy of the high density approximation may be tested in 1-dimension by comparison with exact results. The exact and MF results for the zero temperature longitudinal magnetization of the transverse field Ising chain
\begin{align}
\mathcal{H} = -J_{nn} \sum_i \tau_i^z \tau_{i+1}^z - \Delta_0 \sum_i \tau_i^x
\end{align}
are given by \cite{Pfeuty}:
\begin{align}
\langle \tau^z \rangle &= \biggr(1-(\Delta_0/J_{nn})^2\biggr)^{1/8}
\\ \nonumber 
\langle \tau^z \rangle \biggr|_{MF} &= \biggr(1-(\Delta_0/2J_{nn})^2\biggr)^{1/2}.
\end{align}
The effects of fluctuations about the MF are quite substantial in 1-dimension. We see that MF theory overestimates the critical transverse field by a factor of two, as well as predicting the critical exponent $\beta=1/2$ rather than the exact value of $\beta=1/8$. In Fig. \ref{fig:TFIM1D} we compare the exact result for the longitudinal spin polarization of the transverse Ising chain to the MF result and the result of order $1/z_c$ in the high density approximation. The $1/z_c$ result is clearly an improvement over MF theory; however, it falls well short of the exact solution. We expect this to be a worst case scenario for two reasons: (i) corrections due to fluctuations become smaller in higher dimensions, and (ii) the high density approximation is rather poor when $z_c=2$. In the dipole-dipole coupled $LiHoF_4$ crystal, the shape dependent effective coordination number is determined by the zero wavevector component of the dipole wave sum $D_0^{zz}$. In a long cylindrical sample of $LiHoF_4$, using the transverse lattice spacing $a=5.175$ Angstroms as reference, we have $J_D / a^3 \approx 7mK$ and $D_0^{zz} a^3 \approx 11.3$ giving an effective coordination number of $z_c^{eff} \approx 11.3$.

\begin{figure}[htp]
\centering
\includegraphics[width=8.5cm]{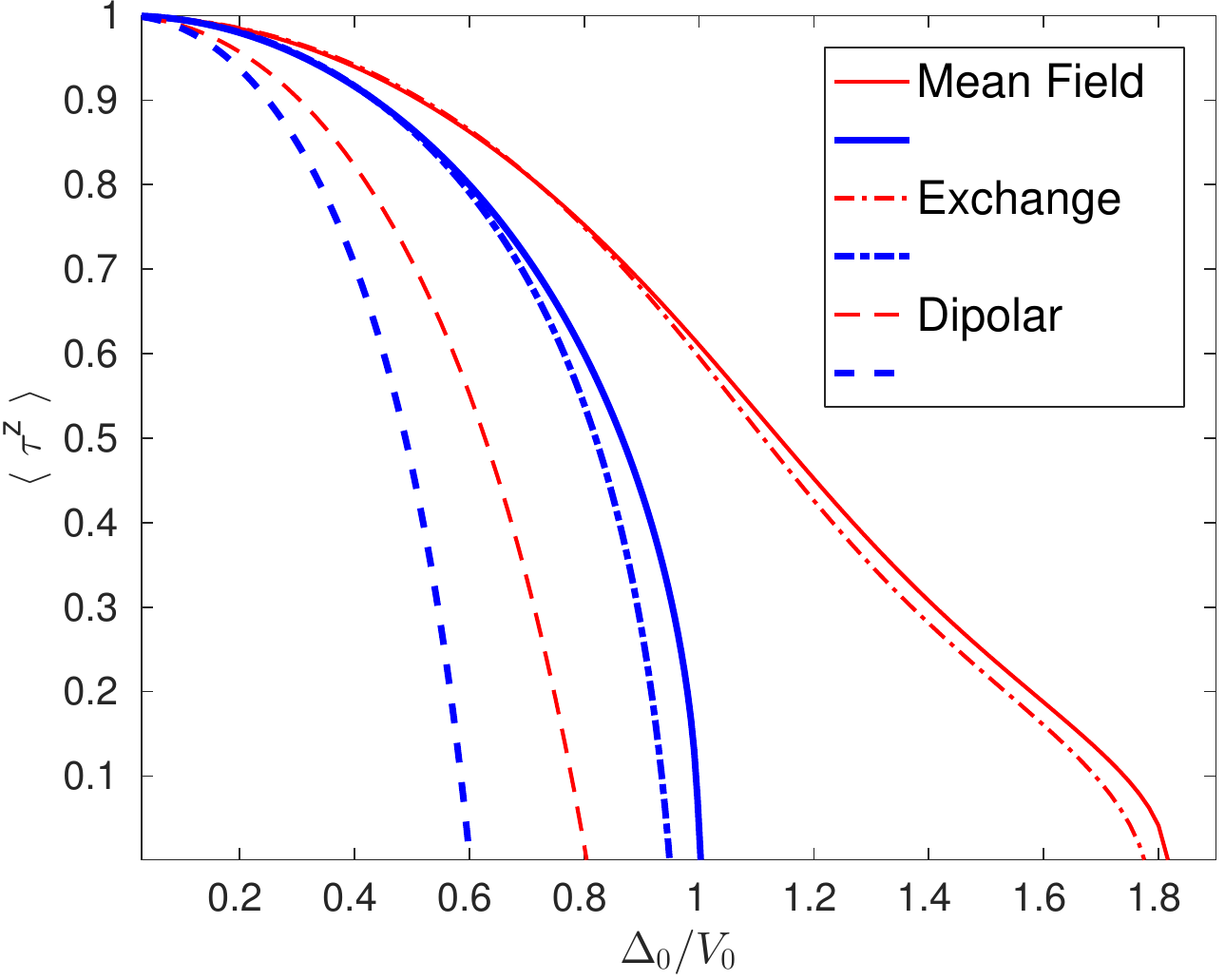}
\caption{Ground state longitudinal spin polarization $\langle \tau^z \rangle$ of the quantum Ising model $\mathcal{H}_0$ in eq. (\ref{eq:H1}) (blue), and the toy model $\mathcal{H}=\mathcal{H}_0+\mathcal{H}_{SB}$ of eq. (\ref{SH2-hyp}) (red) with $A_z = J_{nn}$ and $A_{\perp}=0.1J_{nn}$, as a function of the normalized transverse field $\Delta_0 / V_0$. The MF results (solid line) are for a simple cubic lattice with $V_0 = 6J_{nn}$. The leading order corrections to the MF results are calculated using an expansion in the inverse coordination number $1/z_c$ for both an exchange coupled system $V_{\bf k} = 2J_{nn}(\cos{(k_x)}+\cos{(k_y)}+\cos{(k_z)})$ (dot-dashed line), and a dipolar coupled system $V_{\bf k} = 6J_{D} + 6J_{D}(1-3k_z^2 / {\bf k}^2)$ (dashed line). For the sake of the comparison we take $J_{nn}=J_D$. The lattice spacing is taken to be equal to unity.}
\label{fig:TFIMPolarization}
\end{figure}


In Fig. (\ref{fig:TFIMPolarization}) we show the effect of quantum fluctuations on the longitudinal Ising spin polarization for both the ``bare" quantum Ising model $\mathcal{H}_0$ of eq. (\ref{eq:H1}), and the toy model with added spin bath. Both calculations are done on a 3-dimensional simple cubic lattice. We see that in the models considered here, the spin bath has a substantial quantitative impact on the phase diagram. However, it does not fundamentally change the quantum critical behaviour - we still have a QPT.

In this figure we also show the effect of introducing long-range dipolar interactions between the Ising spins - as occurs in the $LiHoF_4$ system. The effect of quantum fluctuations is now quite striking. This is because for any given spin, the dipole interaction favours anti-alignment of all other spins in the transverse plane, leading to a large enhancement of the quantum fluctuations.

We can now summarize the results of the theory employed here. We have seen that when we include a spin bath in the problem, we can still model the thermodynamic properties using a scalar field theory. In mean field theory, the QPT is not affected by the spin bath, although the critical modes revealed in an RPA analysis now have an ``electronuclear" character, as shown in the spectral weight of the different modes. When we include critical fluctuations, the QPT is not suppressed, with or without a spin bath, although there are corrections to the MFT phase diagram - corrections which are much stronger when the interactions between Ising spins are dipolar. Finally, we see that there is no fundamantal difference between the results for the toy model and for the $LiHoF_4$ system, although obviously there will be quantitative differences. From this point of view we see that the toy model captures the essential behaviour of much more complex systems. 

With all these remarks in mind, it is time to look at the comparison with experiments.


\section{Experiments}
\label{sec:Exp}


Clearly one would like to know how generally applicable are the results derived above, and how one might test for them experimentally. In what follows we do not attempt any kind of complete analysis, but just indicate our main conclusions. 

Let us begin with $LiHoF_4$. We consider the total susceptibility (electronic and nuclear) 
$\chi^{\mu \nu} = -G_{ee}^{\mu\nu}-\gamma (G_{en}^{\mu\nu}+G_{ne}^{\mu\nu})- \gamma^2 G_{nn}^{\mu\nu}$, as discussed in Section \ref{sec:TotalSusc}. In $LiHoF_4$, the nuclear contributions are suppressed by factors of $\gamma = g_n \mu_n / g_L \mu_B \approx 1/550$. The physical transverse spin operators $J^{\mu}$ are linear combinations of the effective spin operators $\tau^{\mu}$, and the corresponding correlation functions are then combinations of the correlation functions of the effective spin operators. Consider as an example $J^y = C_y + C_{yy} \tau^y + C_{yx} \tau^x$. The associated correlation function $G_{ee}^{yy}({\bf k}, \tau) = \langle \delta J_{\bf k}^y(\tau) \delta J_{\bf -k}^y(0)\rangle$ is given by
\begin{align}
G_{ee}^{yy}({\bf k}, \tau) &=  
C_{yy}^2 G_{\tau \tau}^{yy}({\bf k}, \tau)
+ C_{yx}^2 G_{\tau \tau}^{xx}({\bf k}, \tau)
\\ \nonumber
& + C_{yy} C_{yx} 
\bigr(G_{\tau \tau}^{xy}({\bf k}, \tau) +  G_{\tau \tau}^{yx}({\bf k}, \tau)\bigr),
\end{align}
with the $G_{\tau\tau}^{\mu\nu}$ given in (\ref{eq:corr}). Unlike the toy model, $LiHoF_4$ has correlations between the $x, y$, and $z$ components of the effective spin operators, mediated by the crystal electric field. 

\begin{figure}[htp]
\centering
\includegraphics[width=8.5cm]{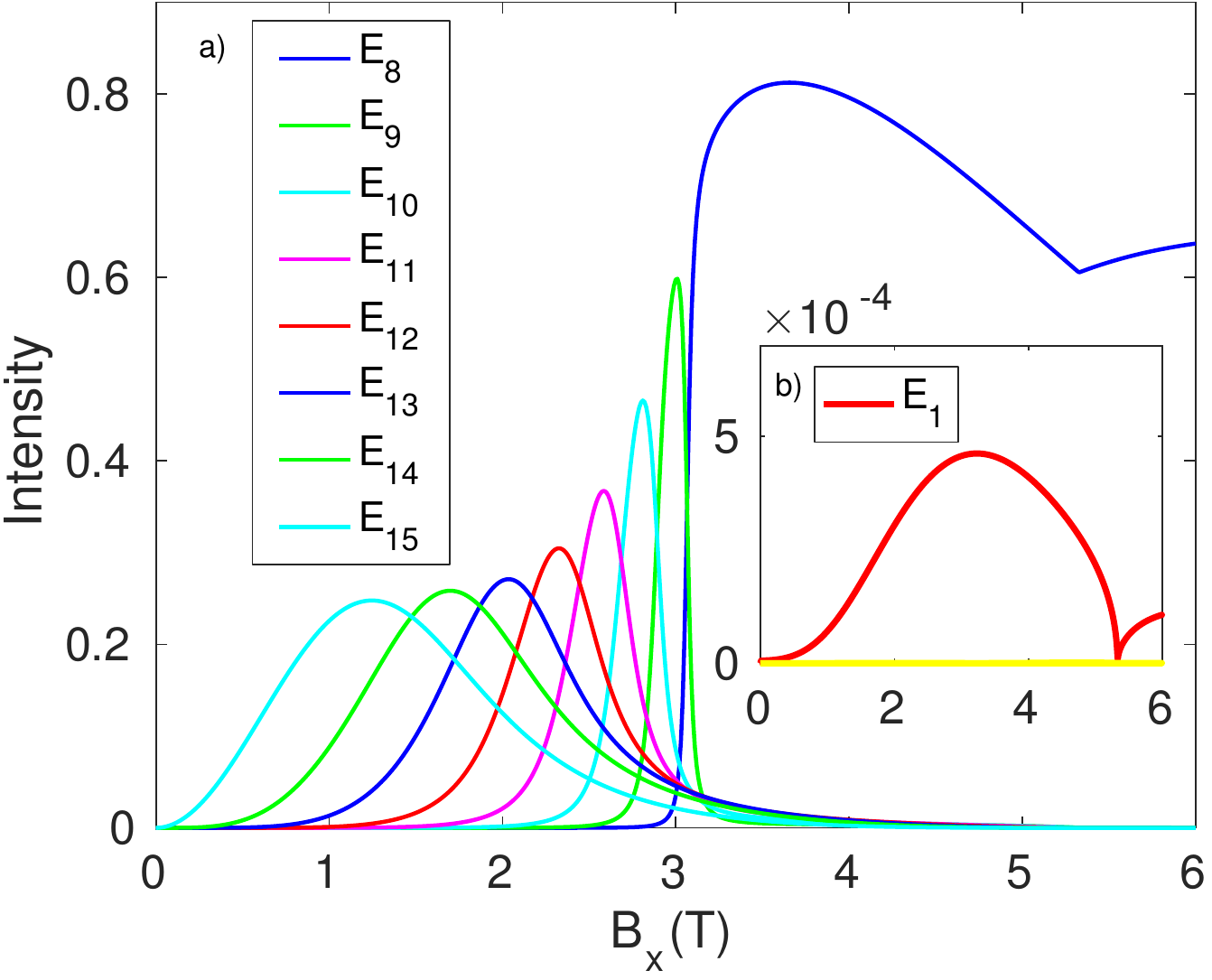}
\caption{Zero temperature intensities $A_p^y$ (arbitrary units), with $\chi_{yy}''(0,\omega) = \sum_n A_p^y \delta(\omega-E_p)$, of the ${\bf k}=0$ RPA modes $\{ E_p \}$ of $LiHoF_4$ associated with the transverse susceptibility, as a function of transverse field $B_x$ (in Tesla). (a) Intensities of the electronuclear modes corresponding to the lowest lying crystal field excitation ($E_8-E_{15}$). (b) The inset shows the intensity of the soft mode (red). The intensities of all other modes (yellow) are negligible.}
\label{fig:SpecWeightLiHoyy}
\end{figure}

In Fig. (\ref{fig:SpecWeightLiHoyy}), we depict the total zero temperature spectral weight of the RPA modes of $LiHoF_4$ expected from the $\chi^{yy}({\bf k}, \omega)$ response. We see that there is very little absorption due to the low energy modes, with the soft mode dominating any absorption that does occur; the $\chi^{xx}({\bf k}, \omega)$ response is similar. At the QPT, the weight of the soft mode seen in $\chi^{xx}$ and $\chi^{yy}$ vanishes, and only the higher lying crystal field excitations are able to absorb energy; however, the soft mode should be visible at the QPT in $\chi^{zz}({\bf k}, \omega)$, as illustrated in Fig. \ref{fig:RPASpectrumLiHo}.


In very recent work \cite{KovacevicRonnow16}, NMR was used to observe the absorption of the low energy electronuclear modes in $LiHoF_4$. In these experiments, the transverse susceptibility $\chi^{yy}({\bf k}, \omega)$ was measured. On the basis of the theory given here, one can make two remarks:

(a) As noted above, the spectral weight of the soft mode in $\chi^{yy}({\bf k}, \omega)$ should vanish near the QPT - thus, to see this mode, measurements should focus on $\chi^{zz}({\bf k}, \omega)$.

(b) Although our results are not directly comparable with experiment, which are performed at finite T, our results include fluctuations about the MF. Note that the fits given by Kovacevic et al. \cite{KovacevicRonnow16} are to MF theory; they do not include quantum fluctuations, and the soft mode at the phase transition is not apparent.

There are other 3-dimensional quantum Ising systems that can be analyzed in the same way as we have done here. Examples are the molecular magnetic system $Fe_8$ \cite{Takahashi11,MorelloFe8,*BurzuriFe8}, and a number of $Mn$-based molecular magnets \cite{[{For Mn$_{12}$ see }][{}] MillisMn12, *delBarcoMn12, [{ For Mn$_3$ see }][{}] HendersonMn3, [{ For Mn$_4$ see }][{}] QuddusiMn4}. We will look at these systems elsewhere. 

One can also look at lower-dimensional systems, where the $1/z_c$ expansion converges more slowly - nevertheless we still expect that our main results should at least give some guidance as to what to expect. A physical realization of a 1-dimensional quantum Ising system, albeit with weak, frustrated antiferromagnetic couplings between chains, is $CoNb_2O_6$ \cite{ColdeaCoNb, KinrossCoNb, WuCoNb, *LeeBalentsCoNb}. The low energy modes have been probed via NMR \cite{KinrossCoNb}, and the results used to identify scaling regimes predicted in the 1990s \cite{SachdevBook}. The energy spectrum of $CoNb_2O_6$, measured via neutron scattering \cite{ColdeaCoNb}, is gapped at zero wavevector near the critical point, a fact attributed to the weak interchain couplings that cause the system to order at some $k \neq 0$. 

However, we emphasize that, in line with all the results derived in this paper, hyperfine interactions will certainly lead to low energy electronuclear modes, which will need to be included in any theory of the low-energy spectrum. Given the current experimental energy resolution of neutron scattering, the low energy electronuclear soft mode may be indistinguishable from elastic scattering.


\section{Discussion}


We find that Quantum Ising systems coupled to a spin bath, with hybridized modes between the Ising and bath spin variables, must still have a QPT. This is true even when we take account of quantum fluctuations around the MFT results, and for high-spin bath variables. Nor does an independent dynamics for the bath variables (coming from, eg., the extra field $\mbox{\boldmath $\Delta$}_n({\bf B}_x)$ in the $LiHoF_4$ system) change the result. Although long-range dipolar interactions strongly enhance fluctuations around the MFT/RPA results, they also do not destroy the QPT.

It is then worth asking: under what circumstances - if any - can the spin bath destroy the QPT? One clear case occurs when the bath spins are frozen. In our results, we have assumed thermodynamic equilibrium - but at low $T$, bath spin relaxation times to equilibrium can be much slower than the Ising spin dynamics (this is particularly clear when the bath spins are nuclear spins - in this case we know from NMR measurements that relaxation times can become very long). In this case the bath will act as a random static potential on the Ising spins, giving more complex effects. In interpreting any experiment where the system is swept through the QPT at a finite rate, due attention will have to be paid to this point.

In this paper we have only studied the behaviour at temperature $T=0$. For a proper comparison with experiments on systems like $LiHoF_4$, one needs finite-$T$ results - these will be given in another paper. The purpose of the present paper has been to introduce the main methods and show how they are used, and to resolve one key question, viz., how the spin bath influences the QPT that is found in the standard Quantum Ising system. In the course of this work, we have found that our toy model actually is a very good guide to the behaviour in much more complicated systems like $LiHoF_4$. For this reason it is worth studying in its own right - this will be done in more detail elsewhere.


\section{Acknowledgements}


We thank NSERC in Canada for funding, and G. Aeppli, T. Cox and A. Gomez for discussions.

\begin{appendices}


\section{Spin Cumulants}
\label{ap:Cumulants}

In the field theoretic formalism introduced in Section \ref{sec:FormToy}, cumulants of the longitudinal spin operator $M_n(S^z(\tau_1)\ldots S^z(\tau_n))$ taken with respect to the MF Hamiltonian $\mathcal{H}_{MF}$ at different imaginary time indices $\{\tau_n\}$ play a central role. These cumulants may easily be calculated in terms of parameters in $\mathcal{H}_{MF}$; however, the results are quite lengthy. The complexity of the expressions for the spin cumulants is the primary factor limiting the utility of the effective field formalism. In this appendix, we present explicit expressions for the cumulants of up to four spins.

In order to calculate the cumulants, we make use of single-site Hubbard operators $X_{mn} = |m\rangle\langle n|$, where $|m\rangle$ are the eigenstates of the single-site MF Hamiltonian $\mathcal{H}_{MF}^j$ \cite{Hubbard,Yang}; as in the main text, we define matrix elements $c_{mn} = \langle m| \tau^z|n \rangle_0$ between these states. 

In terms of the Hubbard operators the single-site MF Hamiltonian is given by $\mathcal{H}_{MF}^j = \sum_n E_n X_{nn}$ and the longitudinal spin operator is given by
\begin{align}
S^z = \sum_n c_{nn} X_{nn} + \sum_{m \neq n} c_{mn} X_{mn}.
\end{align}
For a single spin we find
\begin{align}
M_1(S^z) = \langle S^z \rangle_0 = \sum_n c_{nn} p_n,
\end{align}
where $p_n = e^{-\beta E_n} / Z_{MF}$ gives the population of the $nth$ eigenstate, and the MF partition function is $Z_{MF}=\sum_n e^{-\beta E_n}$.

Defining $S_i^z \equiv S^z(\tau_i)$ and $X_{mn}^i \equiv X_{mn}(\tau_i)$, where the imaginary time dependence of the operators is given by $O(\tau) = e^{\tau \mathcal{H}_{MF}} O e^{-\tau \mathcal{H}_{MF}}$, we find the two spin cumulant $M_2(S_1^z S_2^z) = \langle S_1^z S_2^z \rangle_0 - \langle S_1^z \rangle_0 \langle S_2^z \rangle_0$ to be
\begin{align}
M_2(&T_{\tau} S_1^z S_2^z)
= \sum_{m,n}  c_{mm}c_{nn} \langle T_\tau X_{mm}^1 X_{nn}^2 \rangle_0
\\ \nonumber
&+ \sum_{P\{1,2\}} \sum_{n > m} c_{mn}^2 \langle T_\tau X_{mn}^1X_{nm}^2 \rangle_0 - \biggr[\sum_m c_{mm} p_m\biggr]^2
\end{align}
where $P\{i\}$ denotes the set of all permutations. The imaginary time ordered products of the Hubbard operators are
\begin{align}
-\langle T_{\tau} X_{nm}(\tau')X_{mn}(\tau) \rangle = p_{mn}K_{mn}^{0}(\tau'-\tau),
\end{align}
with $p_{mn} = p_m - p_n$ being thermal factors, and (in Matsubara frequency space)
\begin{align}
\label{eq:Kmn}
K_{mn}^{0}(i \omega_{r})
= \int_{0}^{\beta} d\tau K_{mn}^0(\tau) e^{i\omega_{r} \tau}
= \frac{1}{E_{mn}-i\omega_{r}},
\end{align}
where $E_{mn}=E_m-E_n$ is the energy difference between the $m^{th}$ and $n^{th}$ MF eigenstate. 

To find the cumulants of three or more spins we employ a general reduction scheme \cite{Yang} for the Hubbard operators
\begin{align}
\langle T_{\tau} &O(\tau_1) \cdots X_{mn}(\tau) \cdots O(\tau_i) \rangle_0 =
\\ \nonumber
& K_{mn}^0(\tau_1-\tau) \langle T_{\tau} [O(\tau_1),X_{mn}(\tau_1)] \cdots O(\tau_i) \rangle_0  \\ \nonumber
&+ K_{mn}^0(\tau_2-\tau) \langle T_{\tau} O(\tau_1)[O(\tau_2),X_{mn}(\tau_2)] \cdots O(\tau_i) \rangle_0
\\ \nonumber
&\cdots
+ K_{mn}^0(\tau_i-\tau) \langle T_{\tau} O(\tau_1) \cdots [O(\tau_i),X_{mn}(\tau_i)] \rangle_0,
\end{align}
where $O(\tau) = X_{pq}(\tau)$ denotes an arbitrary Hubbard operator, and $K_{mn}^0(\tau'-\tau)$ is given in (\ref{eq:Kmn}). The $n^{th}$ spin cumulant may be written as
\begin{align}
\label{eq:M-rec}
M_n(S_1^z \ldots &S_n^z) = \langle S_1^z \ldots S_n^z \rangle_0
\\ \nonumber
&- \sum_{n_1+\ldots+n_k =n}M_{n_1}M_{n_2} \ldots M_{n_k}, 
\end{align}
so at each order the only new term that needs to be computed is the $n$ spin correlation function $\langle S_1^z \ldots S_n^z \rangle_0$. In the quantum ($T=0$) regime, terms in the $n$ spin correlation function cancel with all the lower order cumulants leading to a significant simplification. Here, we simply give the final results in the limit $T \rightarrow 0$ since they are necessary for the evaluation of the fluctuation corrections contained in Section \ref{sec:Qfluc}. In order to present the results, we define the following functions
\begin{align}
\label{eq:K-def}
K_{mn}^{pq}(i\omega_{r};i\omega_{s}) &\equiv K_{mn}^0(i\omega_{r})K_{pq}^0(i\omega_{s})
\\ \nonumber
^{rs}K_{mn}^{pq}(i\omega_{r};i\omega_{s};i\omega_q) &\equiv K_{mn}^0(i\omega_r)K_{pq}^0(i\omega_s)K_{rs}^0(i\omega_q), 
\end{align}
which simply represent chains of propagators between MF eigenstates. For brevity, we make use of the notation $f(i\omega_{r_1} \ldots i\omega_{r_n}) = f(r_1 \ldots r_n)$ in the functions used below, in discussing 3rd-order and higher cumulants.

\vspace{2mm}

{\bf 2nd-order Cumulant:} Contracting the Hubbard operators and transforming to Matsubara frequency space
\begin{align}
M_2(i\omega_{r_1},i\omega_{r_2})
= \prod_{i=1}^2 \int_0^{\beta} d\tau_i e^{i\omega_{r_i}\tau_i} M_2(T_{\tau} S_1 S_2),
\end{align}
we find the second order cumulant to be given by
\begin{align}
\label{eq:M2}
M_2(i&\omega_{r_1},i\omega_{r_2}) = \beta \sum_{n > m}
\frac{2 c_{mn}^2 p_{mn} E_{nm}}{E_{nm}^2-(i\omega_{r_1})^2}
\delta_{\omega_{r_1}+\omega_{r_2},0}
\\ \nonumber
&+\beta^2 \biggr(\sum_{m}c_{mm}^2p_m - \biggr[\sum_m c_{mm}p_m \biggr]^2\biggr) \prod_{i=1}^2 \delta_{\omega_{r_i},0},
\end{align}
The second term in (\ref{eq:M2}) vanishes in the zero temperature limit, and in the disordered phase of the system.

The connected MF longitudinal imaginary time correlation function $g(\tau) = -\langle \delta S^z(\tau) \delta S^z(0)\rangle_0$ follows from the two spin cumulant via a frequency summation $g(i \omega_{r_1}) = -\beta^{-1} \sum_{r_2} M_2(\omega_{r_1},\omega_{r_2})$. Performing the frequency summation we find
\begin{align}
g(i\omega_{r}) &=
- \sum_{n > m}c_{mn}^2 p_{mn} \frac{2 E_{nm}}{E_{nm}^2-(i\omega_{r})^2}
\\ \nonumber
&-\beta \biggr(\sum_{m}c_{mm}^2p_m - \biggr[\sum_m c_{mm}p_m \biggr]^2\biggr) \delta_{\omega_{r},0}.
\end{align}
which we use in the discussion of RPA correlators in the main text.

\vspace{2mm}

{\bf 3rd-order Cumulant}: In the low temperature limit, the third order spin cumulant is found from eqtns. (\ref{eq:M-rec}) and (\ref{eq:K-def}) to be
\begin{align}
\label{eq:M3}
\lim_{T \rightarrow 0} M_3(r_1,&r_2,r_3)
\;=\; \sum_{n>1} (c_{11}-c_{nn}) |c_{1n}|^2 A_{n}^0(r_1,r_2,r_3)
 \nonumber \\
&+ \sum_{\stackrel{n>1}{p>n}} \text{Re}[c_{1n} c_{np} c_{p1}] A_{np}^0(r_1,r_2,r_3),
\end{align}
where we use the abbreviated notation described above for the the Matsubara frequencies, and where
\begin{align}
\label{eq:A0np}
A_{n}^0(r_1,&r_2,r_3) = \beta \sum_{P\{\omega_{r_i}\}}
K_{n1}^{1n}(\omega_{r_1};\omega_{r_2}) \delta_{\sum \omega_{r_i},0}
\\ \nonumber
A_{np}^0(r_1,&r_2,r_3) =
\\ \nonumber
&2\beta \sum_{P\{\omega_{r_i}\}}
\frac{E_{p1}E_{n1}-(i\omega_{r_1})(i\omega_{r_2})}{(E_{p1}^2-(i\omega_{r_1})^2)
(E_{n1}^2-(i\omega_{r_2})^2)}  \delta_{\sum \omega_{r_i},0}.
\end{align}
in which as before we have defined $p_{mn} = p_m - p_n$, with $p_n = Z_{MF}^{-1} e^{-\beta E_n}$. The notation $\sum_{{P\{\omega_{r_i}\}}}$ indicates a sum is to be performed over every permutation of the Matsubara frequencies. This result for the three spin cumulant is necessary for calculating the leading order corrections to the mean field magnetization in the high density approximation.

\vspace{2mm}

{\bf (ii) 4th-order Cumulant}: In the low temperature limit, the same techniques give the fourth order cumulant as
\begin{align}
\label{eq:M4}
\lim_{T \rightarrow 0} &M_4(\{ i\omega_{r_i} \})
= \sum_{m \neq n} c_{mm}^2|c_{mn}|^2 B_1
\\ \nonumber
&+ \sum_{n > m} c_{mm}c_{nn}|c_{mn}|^2 B_2
+ \sum_{n > m} |c_{mn}|^4 B_3
\\ \nonumber
&+\sum_{m \neq n \neq p} c_{mm}c_{mn}c_{np}c_{pm} B_4
+\sum_{p>n>m} |c_{mn}|^2|c_{mp}|^2 B_5
\\ \nonumber
&+ \sum_m \sum_{n>m} \sum_{\stackrel{p>m}{p\neq n}} \sum_{\stackrel{q>m}{q\neq n,p}}
c_{mn}c_{np}c_{pq}c_{qm} B_6,
\end{align}
where the coefficients $B_1 - B_6$ are defined in terms of the functions in (\ref{eq:K-def}) as
\begin{align}
\nonumber
B_1
=&\beta p_{mn}\sum_{P\{\omega_{r_i}\}}
\ ^{nm}K_{mn}^{mn}(r_1;r_1+r_2;r_3)
\delta_{\sum \omega_{r_i},0}
\end{align}
\begin{align}
\nonumber
B_2
=&-2\beta p_{mn}\sum_{P\{\omega_{r_i}\}}
\ ^{mn}K_{nm}^{mn}(r_1;r_2+r_3;r_2)
\delta_{\sum \omega_{r_i},0}
\end{align}
\begin{align}
\nonumber
B_3
=&\beta p_{mn} \sum_{P\{\omega_{r_i}\}}
\ ^{mn}K_{nm}^{nm}(r_1;r_2;r_3) \delta_{\sum \omega_{r_i},0}
\end{align}
\begin{align}
\nonumber
B_4 = & \sum_{P\{\omega_{r_i}\}} \beta\biggr(p_{mn}
\ ^{mn}K_{np}^{mn}(r_1;r_2;r_2+r_3)
\\ \nonumber
&\ \ + p_{mp} \ ^{pm}K_{np}^{mp}(r_1;r_1+r_2;r_3)
\biggr)\delta_{\sum_{\omega_{r_i}},0}
\end{align}
\begin{align}
\nonumber
B_5 =& \beta \sum_{P\{\omega_{r_i}\}} \biggr[
p_{mp}\ ^{mp}K_{nm}^{pm}(r_1;r_2;r_3)\delta_{\sum \omega_{r_i},0}
\\ \nonumber
&\ \ + p_{mn}\ ^{mn}K_{nm}^{pm}(r_1;r_2;r_3)\delta_{\sum \omega_{r_i},0}-
\\ \nonumber
&\ \ -p_{np} \ ^{np}K_{nm}^{pm}(r_1;r_2;r_1+r_3)
\delta_{\sum \omega_{r_i},0}\biggr]
\end{align}
\begin{align}
\nonumber
B_6 =&\beta \sum_{P\{\omega_{r_i}\}} \biggr[
p_{mq}\ ^{mq}K_{nm}^{np}(r_1;r_2+r_3;r_2)
\delta_{\sum \omega_{r_i},0}
\\ \nonumber
&\ \ + p_{qp}\biggr(\ ^{qp}K_{nm}^{mp}(r_1;r_2+r_3;r_2)
\\ \nonumber
&\ \ -\ ^{qp}K_{nm}^{pn}(r_1;r_2;r_3)\biggr)\delta_{\sum \omega_{r_i},0}+
\\
&\ \ + p_{qn}\ ^{qn}K_{nm}^{pn}(r_1;r_2;r_2+r_3)
\delta_{\sum \omega_{r_i},0}\biggr].
\end{align}

Note that these results are quite general - in the main body of the paper we have used them for the toy model and for the $LiHoF_4$ system. The four spin cumulant appears in the quartic term $u_4$ of the auxiliary field theory. We require its zero frequency and wavevector limit in order to determine the stability of the theory.

\section{Magnetization Corrections}
\label{ap:MagCorrections}

Here we express the leading order correction to the MF magnetization (see Sections \ref{sec:FormToy} and \ref{sec:Qfluc})
\begin{align}
\langle \delta S_{\bf k}^z(\tau)\rangle
= \frac{1}{\sqrt{\beta U_{\bf k}}} \biggr\langle \phi_{\bf k}(\tau) \biggr\rangle_{\phi}
\end{align}
in terms of parameters in $\mathcal{H}_{MF}$. The average on the right $\langle \cdots \rangle_{\phi}$ is determined with respect to the effective Hamiltonian for the auxiliary field $\mathcal{H}_{eff}[\phi]$ given in (\ref{eq:Heff}). We perform perturbation theory in the mode-mode interactions between the auxiliary field fluctuations. As discussed in Section \ref{sec:FlucEffects}, this leads to and expansion in the inverse coordination number $1/z_c$, with the leading order correction to the MF magnetization involving a single power of $u_3$. Writing $\langle \delta S_{\bf k}^z(\tau)\rangle = \langle S_{\bf k}^z(\tau)\rangle_1 + \langle S_{\bf k}^z(\tau)\rangle_2 + \cdots$, and transforming to Matsubara frequency space ($S_{\bf k}^z(\tau) = \beta^{-1} \sum_r e^{-i\omega_r \tau} S_{\bf k}^z(i\omega_r)$), we find
\begin{align}
\langle S_{\bf k}^z(i\omega_r) \rangle_1 &= \frac{-1}{\sqrt{\beta V_{\bf k}}}
\sum_{\{r_i , {\bf k}_i\}} \frac{u_3}{3!}
\biggr\langle \phi_{\bf k}^r \phi_{{\bf k}_1}^{r_1}\phi_{{\bf k}_2}^{r_2}\phi_{{\bf k}_3}^{r_3}
\biggr\rangle_{\phi},
\end{align}
where we define $\phi_{\bf k}^{r} \equiv \phi_{\bf k}(i\omega_{r})$ for brevity. This correction is of order $1/z_c$ in the high density approximation, with the next contribution involving both $u_3$ and $u_4$ being of order $1/z_c^3$. Contracting the fields and making use of the explicit expression for $u_3$ given in (\ref{eq:un}) we find
\begin{align}
\label{eq:mag}
\langle S_{\bf k}^z(i\omega_r) \rangle_1 =
-\frac{\mathcal{D}_{\bf k}^o(i\omega_r)}{2\beta^2\sqrt{N}}
\sum_{r',{\bf k}'} {\cal T}_{{\bf k}'}(i\omega_{r'})
M_3(r,r',-r'),
\end{align}
with $M_3(r_1,r_2,r_3) \equiv M_3(i\omega_{r_1},i\omega_{r_2},i\omega_{r_3})$, the third order cumulant, given in Appendix \ref{ap:Cumulants}. Note that a factor of one half has been introduced into (\ref{eq:mag}) to account for the fact that the integration over the fields double counts each degree of freedom. The renormalized interaction between the spins, or T-matrix, is defined by
$\mathcal{T}_{\bf k}(i\omega_{r}) = V_{\bf k} \mathcal{D}_{\bf k}^o(i\omega_r)$. Quite generally, contractions of the field operators in the perturbation expansion lead to a renormalization of the powers of the interaction appearing in the $\{u_n\}$. Put another way, magnetic fluctuations renormalize the bare interaction between spins.

The static correction to the magnetization for a spin at a single site follows from $\langle S^z \rangle_1 = \langle S_{{\bf k}=0}^z(i\omega_r=0) \rangle_1 / \sqrt{N}$; it follows that
\begin{align}
\langle S^z \rangle_1 = -\frac{R_0}{2N \beta^2} \sum_{r,{\bf k}}  {\cal T}_{\bf k}(i\omega_{r})
\;  M_3(0,i\omega_{r},-i\omega_{r}),
\end{align}
where $R_0 = \mathcal{D}_{{\bf k}=0}^o(0)$ is the zero frequency and wavevector component of the free field propagator. We restrict our attention to the quantum ($T=0$) limit, in which case the T-matrix is
\begin{align}
{\cal T}_{\bf k}(i\omega_r)\biggr|_{T=0} = V_{\bf k} \frac{\prod_{n>1} (E_{n1}^2-(i\omega_r)^2)}{\prod_p((E_{\bf k}^p)^2-(i\omega_r)^2)},
\end{align}
where the $E_{nm}$ are energy differences between MF eigenstates, and the $E_{\bf k}^p$ are the energies of the RPA modes of the system. At zero temperature, $R_0$ is simply the ratio
\begin{align}
R_0 \biggr|_{T=0} =\frac{\prod_{n>1} E_{n1}^2}{\prod_p (E_{{\bf k}=0}^p)^2}.
\end{align}

Combining terms, and incorporating (\ref{eq:M3}) and (\ref{eq:A0np}), we find the leading-order quantum correction to the MF magnetization to be
\begin{align}
\label{eq:Sz1}
\langle S^z \rangle_1 = \frac{R_0}{2N} \sum_{\bf k} V_{\bf k}
\biggr[ &\sum_{n>1} (c_{11}-c_{nn}) |c_{1n}|^2 \chi^n
\\ \nonumber
&+ \sum_{p>n\neq 1} \text{Re}[c_{1n}c_{np}c_{p1}] \chi^{np}   \biggr],
\end{align}
where the $c_{mn}$ are the MF matrix elements of the longitudinal spin operator, and
\begin{widetext}
\begin{align}
\chi^n &= -\frac{2}{\beta}\sum_{r}
\frac{3E_{n1}^2-(i\omega_{r})^2}{E_{n1}^2-(i\omega_{r})^2}
\frac{\prod_{m\neq n,1} (E_{m1}^2-(i\omega_{r})^2)}
{\prod_p ((E_{\bf k}^p)^2-(i\omega_{r})^2)}
\\ \nonumber
\chi^{np} &= \frac{4}{\beta} \sum_{r} \frac{1}{\prod_{l}(E_{\bf k}^l)^2-(i\omega_{r})^2} \biggr[\frac{E_{n1}}{E_{p1}} \prod_{m \neq n,1} (E_{m1}^2-(i\omega_{r})^2)
\\ \nonumber
& \qquad\qquad\qquad\qquad
+ \frac{E_{p1}}{E_{n1}} \prod_{m \neq p,1} (E_{m1}^2-(i\omega_{r})^2)
+ (E_{p1}E_{n1}+(i\omega_{r})^2) \prod_{m \neq n,p,1} (E_{m1}^2-(i\omega_{r})^2)\biggr].
\end{align}
To find the magnetization corrections given in the main text we need to perform the frequency summations in $\chi^n$ and $\chi^{np}$. For $\chi^n$ the result is
\begin{align}
\label{eq:Xn}
\chi^n|_{E_n \neq E_{\bf k}^p} = -2\biggr[E_{n1}&
\frac{\prod_{m\neq n,1} (E_{m1}^2-E_{n1}^2)}{\prod_p ((E_{\bf k}^p)^2-E_{n1}^2)}
\coth{(\frac{\beta E_{n1}}{2})}
\\ \nonumber
&+\sum_{p}\frac{3E_{n1}^2-(E_{\bf k}^p)^2}{E_{n1}^2-(E_{\bf k}^p)^2}
\frac{\prod_{m\neq n,1} E_{m1}^2-(E_{\bf k}^p)^2}{2E_{\bf k}^p\prod_{q\neq p} (E_{\bf k}^q)^2-(E_{\bf k}^p)^2}\coth{(\frac{\beta E_{\bf k}^p}{2})}\biggr],
\end{align}
provided that none of the differences between MF energy levels are degenerate with the energies of the RPA modes; if a MF level is degenerate with an RPA mode, we simply shift the MF energy level by a small amount to avoid dealing with a higher order pole. The $\chi^{np}$ term yields
\begin{align}
\label{eq:Xnp}
\chi^{np} = \sum_{l} \frac{2}{E_{\bf k}^l\prod_{q \neq l}(E_{\bf k}^q)^2-(E_{\bf k}^l)^2} &\biggr[
\frac{E_{n1}}{E_{p1}} \prod_{m \neq n,1} (E_{m1}^2-(E_{\bf k}^l)^2) + \frac{E_{p1}}{E_{n1}} \prod_{m \neq p,1} (E_{m1}^2-(E_{\bf k}^l)^2)
\\ \nonumber
&+ (E_{p1}E_{n1}-(E_{\bf k}^l)^2) \prod_{m \neq n,p,1} (E_{m1}^2-(E_{\bf k}^l)^2)\biggr]
\coth{(\frac{\beta E_{\bf k}^l}{2})}.
\end{align}
\end{widetext}

In the zero temperature limit, $\coth{(\frac{\beta E_{\bf k}^l}{2})}$ is simply equal to one. In the case of a spin half transverse Ising system without a spin bath (a two level system with a single RPA mode), all the fluctuation results reduce to those of Stinchcombe \cite{Stinchcombe1,Stinchcombe2,Stinchcombe3}, which we have derived here in a new way. Equations (\ref{eq:Sz1}), (\ref{eq:Xn}) and (\ref{eq:Xnp}), are used to obtain the results presented in Section \ref{sec:Qfluc}.

\end{appendices}

\bibliography{PRBbiblio}

\end{document}